\documentclass[letterpaper]{article} 
\usepackage{aaai24}  
\usepackage{times}  
\usepackage{helvet}  
\usepackage{courier}  
\usepackage[hyphens]{url}  
\usepackage{graphicx} 
\usepackage{natbib}  
\usepackage{caption} 
\usepackage{xcolor}
\usepackage{booktabs}

\title{Surveys Considered Harmful? \\Reflecting on the Use of Surveys in AI Research, Development, and Governance}
\author {
    Mohammad Tahaei\textsuperscript{\rm 1, 2},
    Daricia Wilkinson\textsuperscript{\rm 3},
    Alisa Frik\textsuperscript{\rm 1},
    Michael Muller\textsuperscript{\rm 4},
    Ruba Abu-Salma\textsuperscript{\rm 5},
    Lauren Wilcox\textsuperscript{\rm 2}
}
\affiliations {
    \textsuperscript{\rm 1}International Computer Science Institute,  \textsuperscript{\rm 2}eBay, 
    \textsuperscript{\rm 3}Arizona State University, 
    \textsuperscript{\rm 4}IBM Research,
    \textsuperscript{\rm 5}King's College London\\
    mtahaei@icsi.berkeley.edu, daricia.wilkinson@asu.edu, afrik@icsi.berkeley.edu, michael\_muller@us.ibm.com,
    ruba.abu-salma@kcl.ac.uk, lgw231@acm.org}

\begin{document}

\maketitle

\begin{abstract}
Calls for engagement with the public in Artificial Intelligence (AI) research, development, and governance are increasing, leading to the use of surveys to capture people's values, perceptions, and experiences related to AI. In this paper, we critically examine the state of human participant surveys associated with these topics. Through both a reflexive analysis of a survey pilot spanning six countries and a systematic literature review of 44 papers featuring public surveys related to AI, we explore prominent perspectives and methodological nuances associated with surveys to date. We find that public surveys on AI topics are vulnerable to specific Western knowledge, values, and assumptions in their design, including in their positioning of ethical concepts and societal values, lack sufficient critical discourse surrounding deployment strategies, and demonstrate inconsistent forms of transparency in their reporting. Based on our findings, we distill provocations and heuristic questions for our community, to recognize the limitations of surveys for meeting the goals of engagement, and to cultivate shared principles to design, deploy, and interpret surveys cautiously and responsibly.
\end{abstract}

\section{Introduction}
\label{sec:introduction}

Artificial Intelligence (AI)\footnote{We use the term ``\textit{AI}'' broadly in our Introduction and Discussion, adopting the US National Institute for Standards and Technology (NIST) definition of the term, as ``\textit{an engineered or machine-based system that
can, for a given set of objectives, generate outputs such as predictions, recommendations, or decisions influencing real or virtual environments. AI systems are designed to operate with varying levels of autonomy}''~\cite{standards2023ai}. We acknowledge that its meaning and scope remain in flux. Our paper includes a systematic literature review, so we rely on authors' definitions of AI when presenting analyses of their work.} and Machine Learning (ML) researchers, developers, and policymakers are increasingly using surveys to capture people's values, perceptions, and experiences, to inform development and governance of AI. Surveys are used to guide the design and development of new technology directions and products \cite[e.g.,][]{alkhathlan2024balancing, sindermann2021assessing, persson2021we, othman2023understanding, loefflad2024types, davani2024disentangling}, shape companies' technology  policies~\cite{anthropic2023,google2024global,openai2023officialsurvey}, and inform national and international policies~\cite{ada2023UK, govai2023NAIAC, un2024AI}. However, critical perspectives caution that if human participant research methods are poorly designed or applied (e.g., embed biases or lack context), they may fail to serve their intended purpose, possibly leading to ethics and participation washing, and other forms of harm, instead of being beneficial~\cite{cooper2022systematic, groves2023goingpublicpaper}.

In this paper, we critically examine the use of surveys in AI research, development, and governance, as they are recurringly used to assess people's subjective views and experiences of AI~\cite{van_berkel2023methodology}. Surveys, and related research instruments such as questionnaires, inherently employ abstraction and reduction as methods of knowing and understanding~\cite{ornstein2013companion}, which may result in overlooking nuances that on the surface level may seem subtle, but in practice can result in amplifying biases and leading to harms~\cite{bhopal2004review, proctor2008agnotology, roberts2012fatal}. The potential misrepresentation of marginalized perspectives by surveys, though unexplored in the AI domain, has been evident in other fields~\cite{mir2012principles, nazroo2007black, nierkens2006smoking,agyemang2009cardiovascular}. For example, in the United Kingdom, ``\textit{nationally-representative}''\footnote{Throughout this paper, we refer to ``\textit{representation}'', ``\textit{representative},'' or ``\textit{nationally-representative}'' as it is used in the cited resources. The notion of representation in ML has been examined by \citet{chasalow2021representativeness}, and we will also address limitations of representation in surveys in our discussion.} surveys measuring tobacco and alcohol use showed significant discrepancies in data collected from minorities~\cite{bhopal2004review}. Results varied substantially for marginalized ethnic groups between different survey agencies, e.g., one set of results reporting a 1\% smoking rate in Bangladeshi women, and another reporting 6\% in the same group, a discrepancy not observed in the majority group self-identifying as European~\cite{bhopal2004review}. Other researchers also showed that commercial and government entities have postponed or prevented action on critical public health matters for vulnerable groups as a result of poor survey research practices~\cite{proctor2008agnotology}. 
In parallel, researchers criticize the focus of AI research on Western, Educated, Industrial, Rich, and Democratic (WEIRD) populations, arguing that it may not accurately represent the experiences and concerns of diverse global populations affected by or interacting with AI~\cite{septiandri2023weird, van_berkel2023methodology}.

Despite these limitations, the use of surveys is expanding rapidly to capture values and normative expectations, and monitor AI-related impacts on people~\cite[e.g.,][]{jakesch2022how,scharowski2023certification, ribeiro2019microtargeting, kramer2018when, arai2023public, bartneck2023personality, ikkatai2023segmentation, yigitcanlar2020contributions}. This pattern follows a trend that began in computing research in the 1980s~\cite{fowler2013survey}. Surveys now directly shape a number of high-stakes AI projects, including finance~\cite{hertzberg2010information}, employment~\cite{franken2019impact,houser2019can}, smart healthcare~\cite{morley2020ethics,sunarti2021artificial}, transportation~\cite{bharadiya2023artificial}, and education~\cite{blodgett2021risks,zanetti2019psychopathic, lunich2024explainable}. They are also used by companies creating the most popular generative models---often with the embedded assumption that they can capture diverse perspectives and the contextual and cultural specificity of the subject matter, e.g., Collective Intelligence Project~\cite{huang2024collective}, Our Life with AI~\cite{google2024global}, and the Moral Machine experiment, which is \textit{``developing global, socially acceptable principles for machine ethics.''}~\cite{awad2018moralmachine}

This paper contributes to the research challenging the use of decontextualized, unidirectional human participant research methods merely as a means to justify technological advancement~\cite{cooper2022systematic, groves2023goingpublicpaper, sloan2022participation}. It aligns with a broader call within the sociotechnical research community to scrutinize research methods, processes, and practices, not just artifacts or outcomes~\cite{mann2019big, irani2010postcolonial, dourish2020on, ali2016does, cooper2022systematic}. Our goal is to guide people using such methods toward a critical examination of AI-related survey design processes and data collection, analysis, and interpretation methods---toward more equitable and just research practices that respect, rather than misrepresent or exploit, surveyed communities. 

The research questions (RQs) motivating our work are: \textbf{RQ1.} How has research in the relevant body of literature positioned surveys as a way of understanding human values, perceptions, and experiences with regard to AI? What are the elements reported in the literature, and what are the cultural and methodological implications of foregrounding those elements? \textbf{RQ2.} Extending the body of knowledge related to survey methods and epistemology, what are the unique questions that could guide the ethical design, deployment, interpretation, and reporting of (large-scale) survey research on AI topics with human participants, at the particular intersection of survey methodology, AI, and society?

We integrate a \textbf{reflexive analysis} of an international survey pilot with a \textbf{systematic literature review}, to critically examine our assumptions and offer a set of provocations that are vital for the AI research community, as public surveys gain an increasingly stronger foothold in the field. We walk through the design and testing of a pilot survey we conducted to reflect on design decisions and findings with respect to key provocations, and complement this analysis with a review of the methods of 44 survey research papers. 

Papers in our corpus aimed to include large sample sizes ($median=607$ participants). Although 14 out of 44 reviewed papers claimed to have representative samples, inconsistent use of the term ``\textit{representation}'' created an illusion of representation rather than engaging with representation in a meaningful way. This is concerning, as misrepresentation could harm marginalized communities, limit our understanding of such groups, and perpetrate incorrect narratives about AI. Only six papers included authors from the Global South (as per the definition of~\citet{uni2024south}), 11 papers lacked authors from the countries where the studies were conducted, and 38 papers lacked feedback from participants during the design stage. We also reflect on our pilot survey's design and provide a set of heuristic questions related to the use of AI in the design and analysis of surveys, the transparency of research platforms (e.g., Prolific), and the potential harms of conducting surveys across different cultures to characterize  perspectives without using culture-sensitive approaches. These questions aim to help the community rethink who controls the data and how it's obtained, for what purpose it is used, and how the results are interpreted and disseminated. 

We argue that approaches to human participant surveys designed to reach the public to inquire about topics related to AI must be critically examined for their role in perpetuating and maintaining the potential to amplify and exacerbate worrisome power dynamics~\cite{baeza_yates2018bias, bloomberg2023humans}. Shifting from unidirectional survey designs to co-creating survey instruments with the impacted communities---ensuring that the surveys are not only \textit{about} people but designed \textit{with} them---could enable researchers to include multiple knowledge systems and account for power dynamics embedded in knowledge production processes~\cite{garcia2021decolonial, bird2020decolonising,kwet2019digial}.

\section{Related Work}
\label{sec:related-work}

\subsection{A Brief History of Surveys: From Agriculture and Military to Computing and AI} 
\label{sec:rw-surveys}

Modern-day survey methods emerged from a long history of societies that sought measurements of their populations through censuses to make plans essential to core governance (e.g., managing food supplies, distributing land, and managing taxation)---dating back to ancient times~\cite{rossi2013handbook, midena2022towards}. As the focus of studies became more specific, such as examining the economic status of households or conducting consumer research, surveys gained increased popularity over traditional small-scale experimental studies~\cite{rossi2013handbook}. Surveys also played a prominent role in the development and study of \textit{psychometrics}, or measuring people's mental activities. Important concepts in psychometrics include correlation, personality scale and psychometric reliability, experimental designs, and increasingly sophisticated statistical analysis \cite{rust2014modern}.
Significantly abused by \textit{racism} and \textit{oppressive} applications in social policies~\cite{reyes2019eugenic, winston2020scientific}, this evolution is also marked by significant developments, including the emergence of literature on questionnaire design, the introduction of standardized scales like the Likert scale for attitude measurement~\cite{groves2009survey, lee2002cultural}, the establishment of state-supported institutes dedicated to survey research, and the development of technology-assisted survey-taking tools~\cite{rossi2013handbook, fowler2013survey}. A notable example of large-scale survey usage is the extensive surveying of American soldiers returning from World War II---which  provided useful norms for design but systemically excluded women~\cite{epstein2013physiological}. Over time, this movement also created a divide in the empirical research community. One school of thought regards surveys as the ``\textit{language of empirical social research},''~\cite{ornstein2013companion} while others criticize the dominance of survey-based scholarly work that is divorced from theory---coined by Mills as ``\textit{abstracted empiricism}''~\cite{mills2023sociological}.

According to Arnstein's ladder of citizen participation~\cite{arnstein1969ladder}, which describes eight levels of citizen involvement in planning processes in the US, attitude surveys and public inquiries belong to the fourth rung of citizen participation, called ``\textit{consultation}.'' In other words, consulting public opinions is a legitimate step toward understanding their perspectives, but it is not sufficient if used alone. Moreover, if the exchange of data is unidirectional, the providers of the data are viewed as ``\textit{sources}'' of information, raising concerns about the extractive versus participatory nature of surveys when engaging with the public~\cite{ADA-Participatory-Data}.~\citet{arnstein1969ladder} encourages the use of more direct citizen participation modes like committees, partnerships, and community engagement that allow people to engage in planning, decision-making, and policy-making more actively (see also work in participatory survey design \cite{smith2004development, tillyard2019new} and in action research \cite{arcaya2018community, hayes2014knowing}).

Despite these criticisms, surveys have become widely used in contemporary research, employed for collecting structured qualitative and statistical data and gaining quantified insights into people' perspectives and attitudes---a prominent methodological cornerstone in areas such as public opinion polls and large-sample approaches to computing research disciplines like HCI and AI more broadly~\cite{fowler2013survey}.\footnote{Human-Computer Interaction (HCI) is referred to as the discipline that provides foundational knowledge for both industry and academic technology research for studying interactions between humans and computers (e.g., from user interaction techniques to sociotechnical systems). User experience is a term commonly used to include, in part, the application of select HCI methods in industry and product settings.} For example, surveys are among the primary methods used to capture user engagement with mass market user interfaces and to provide insights into users' attitudes, experiences, demographics, and psychological characteristics shaping their behavior with technology~\cite{doherty2018engagement,muller2014survey}. Several best practices have been suggested to plan, design, and conduct effective surveys in the real world~\cite{rea2014designing,converse1986survey,dillman1978mail,dillman1998principles,fowler1990standardized,kelley2003good,krosnick1999survey,PewSurveys,tourangeau2000psychology,tourangeau1996asking,Brown2023}, from mail and telephone surveys to web surveys, to help participants comprehend questions, retrieve the information necessary to answer questions, and judge how much information they need to provide~\cite{cannell1977summary}.

Following the trend of using surveys in empirical human participant studies, researchers at AIES and other sociotechnical AI research communities have been using surveys to investigate public perceptions of AI, including the societal risks and expectations associated with AI. For example, \citet{van_berkel2023methodology} found that most papers (published at prominent venues) documenting studies with human participants (65\%, 130 out of 200 papers included in the review) have used surveys to capture perceptions of broadly-defined AI fairness. \textbf{This abundance of survey research underscores the importance of empirically investigating the impact of the use of this method to explore AI-related societal issues~\cite{said2023artificial}.}

\subsection{Rapid Expansion of Survey Use in AI Research, Development, and Governance}
\label{sec:rw-survey-ai-research}

As more empirical and irrefutable evidence emerges, it becomes clear that understanding AI's impact requires a multi-stakeholder effort~\cite{aragon2022human, delgado2023participatory, himmelreich2023against, Torkamaan2024, havens2020situated}. Yet, given the prevailing power asymmetries in this space, AI development is predominantly shaped by industry, research, and policy~\cite{moloi2021artificial}. This dynamic has only recently begun to shift toward public involvement~\cite{birhane2022power, queerinai2023queer, sloan2022participation, dennler2023bound}, though scholars have long advocated for increased public involvement in science and technology as a means to ``\textit{foster greater accountability, better decision outcomes, and increased trust}''~\cite{holdren2011principles, bao2022whose}.

As such, multiple studies have engaged with the public to investigate AI's impact. A plethora of surveys aiming to be representative have been deployed over the years with the goal of examining public awareness of, perceived challenges with, and trust in AI~\cite{zhang2020us} and how the public understands AI~\cite{selwyn2020ai, kieslich2024regulating}, as well as distinguishing between utopian and dystopian narratives surrounding AI~\cite{cave2019scary}.\footnote{Survey methods may have very different uptake among some marginalized populations that have histories of exploitation by quantitatively-oriented majoritarian researchers~\cite{chilisa2019indigenous, denzin2008handbook, kovach2021indigenous, smith2021decolonizing}. Thereby, we urge caution in interpreting claims that a sample has been properly weighted. Even with a \textit{statistical} approach to weighting, discriminatory questions may systematically reduce participation by intimidating some marginalized groups~\cite{berry2020civil}.} A ``\textit{nationally representative}'' (i.e., ``\textit{results are weighted to be representative of the US adult population}''~\cite{zhang2020us}) survey in 2018 with 2,000 Americans showed that most Americans supported AI development but also expressed deep concerns about its future impact~\cite{zhang2020us}. Similarly, a ``nationally representative'' survey (i.e., ``\textit{weighted sample by main demographic characteristics}''~\cite{selwyn2020ai}) with 2019 respondents from Australia supported the development of AI in healthcare but exhibited mixed views on its professional integration~\cite{selwyn2020ai}. These findings resonate with similar survey studies on public perceptions of AI in Russia, India, and the UK~\cite{fast2017long, bao2022whose, kapania2022because,ada2023UK}. Survey results also influence national policies; for example, the frequent citation of survey results in US government policy and strategy documents related to AI~\cite[e.g.,][]{nataird2023strategic, govai2023NAIAC, govai2023strengthening} and surveys' influence on the UK's AI policy~\cite{ada2023UK}. Additionally, periodic public surveys to monitor the evolving AI landscape have been proposed, as evidenced by the National Artificial Intelligence Advisory Committee (NAIAC) and other government documents focused on strengthening AI capabilities~\cite{govai2023NAIAC, govai2023strengthening}. Public surveys are also influencing corporate strategies and future products of companies like Anthropic~\cite{anthropic2023} and OpenAI~\cite{openai2023officialsurvey}.

Recent reflections on participation and inclusion in AI research have contributed to studies extending beyond WEIRD populations to build knowledge representing a broader spectrum of lived experiences~\cite{boyon2022opinions, linxen2021how, epstein2023toward}. For example, \citet{kelley2021exciting}'s survey with 10,000 respondents from eight countries showed their widespread support for AI development and their hesitations because of its associated risks. Participants were hopeful about the potential for AI to improve healthcare, while their fears revolved around concerns over job loss, social isolation, and significant threats to humanity. The data reflected differences in how people in non-Western societies perceived specific risks that media discourses and various predispositional values could shape.

These examples highlight the profound impact that survey research can have on the future of AI and society. Yet, despite the recent surge in the use of surveys, prevailing disciplinary norms and best practices for their design and use for AI research are inadequate.\footnote{There are already well-documented limitations and caveats in survey research to date. Examples include vulnerabilities to biases related to social desirability, order effects, and sampling. In Appendix~\ref{app:known-issues}, we detail some of these known issues with surveys.} As a community, we have yet to discuss essential criteria and summarize principles for the design and use of surveys to understand people's values, perceptions, and experiences with regard to AI. \textbf{We argue that survey research on \textit{AI-related topics} introduces unique methodological challenges and considerations that warrant field-wide attention, further methodological research inquiry, and collaborative debate.}

\subsection{Critically Reflexive Practices in AI Research}
\label{sec:rw-critically-reflexive}
``\textit{Critically reflexive practice embraces subjective understandings of reality as a basis for thinking more critically about the impact of our assumptions, values, and actions on others.}''~\cite{cunliffe2004on} Reflexivity has a history, in scholarly inquiry, of generating  understandings of how underlying scholarly structures and systems influence and are influenced by actors, including scholars themselves~\cite{bourdieu1990other, bourdieu1992invitation, jamieson2023reflexivity}. Despite the recent establishment of sociotechnical AI research communities, a body of work that critically examines the communities' methods and results is rapidly emerging~\cite{groves2023goingpublicpaper,laufer2022four, chasalow2021representativeness, miceli2021documenting,constantinides2024role,constantinides2024method,havens2020situated}. 
\citet{laufer2022four} leaned on reflexivity to provide a nuanced exploration of the ``\textit{presuppositions, underlying values and assumptions}'' guiding the direction of such scholarship since its inception. Similarly, \citet{young2022confronting} reflected on emerging conflicts of interest that inadvertently and inherently impact models of participation with audiences who contend with the negative impacts of algorithmic systems. Researchers have questioned the use of human research methods within the AI industry and their role in driving narratives and redefining norms around empirical research in AI. \citet{groves2023goingpublicpaper} critique the approaches of commercial AI labs, such as OpenAI and Anthropic, in their public engagement strategies. They observe that business interests are often prioritized over societal needs, characterized by a lack of context, clarity in methods, and rigor, stemming from the fast-paced nature of the industry and the conflicting interests involved. Moreover, a reflexive study reveals that the majority of FAccT papers involving human participants predominantly focus on WEIRD populations, particularly those from the US~\cite{septiandri2023weird}. 

Building on the critical self-reflective literature in AI research, our perspective is informed by critical computing~\cite{comber2020announcing, ko2023critically}, a body of work that includes titles with ``\textit{considered harmful}'' (starting in 1968 to challenge research norms and structures~\cite{dijkstra1968harmful}). This resonates across various computing disciplines such as computer security~\cite{singer2021trust} and HCI~\cite{aragon2022human, comber2020announcing, greenberg2008usability, crabtree2009ethnography}. Similarly, the field of critical data studies examines how data are not \textit{given} or even \textit{captured}~\cite{muller2019data}, but rather \textit{designed}~\cite{feinberg2017design, feinberg2022everyday} and \textit{created}~\cite{muller2021designing, muller2022forgetting} as human-made components of larger sociotechnical assemblages of privilege and power~\cite{iliadis2016critical, kitchin2014towards}. \textbf{This paper builds on the trove of existing and emerging research to critically examine the often-overlooked assumptions embedded in the use of surveys within AI scholarship, and to identify opportunities where the AIES community is uniquely positioned to take a leading role.} 

\section{Methods}
Our approach and perspectives are informed by a critically reflexive stance, rooted in the self-critical perspectives of the AI research community (Section~\ref{sec:rw-critically-reflexive}) and our positionality (Section~\ref{sec:positionality}). The rationale behind such a critical stance is to initiate discourse within the community, especially as surveys are increasingly becoming a ``go-to'' method for capturing public perceptions of AI. To understand the pitfalls of using public surveys in the AI domain, we employ two methods: (1) \textbf{a pilot survey as the basis for critical reflection using reflexivity} and (2) \textbf{a systematic literature review of public surveys in AI research}.

\subsection{Pilot Survey of Perceived Benefits and Risks of AI}
\label{sec:pilot-survey}
To facilitate reflexivity in survey methods, we conducted a pilot survey in six countries (one in each of six continents) with 282 participants to explore the complexities of survey research associated with the challenging topic of capturing perceptions of AI's benefits and risks. Despite adhering to \textit{known} survey research best practices (see Appendix~\ref{app:known-issues}), we observed that there are both \textit{unknown knowns} and \textit{unknown unknowns} that require further attention and are often overlooked in current AI research practices.

\textbf{Survey Design.}
 The pilot survey explored perceptions of the \textit{benefits} and \textit{risks} of \textit{existing} AI systems via two separate open-ended questions: (1) How do you think existing AI systems could benefit you?; (2) How do you think existing AI systems could put you at risk? We also used two storytelling questions inspired by the computer security domain~\cite{rader2012stories, pfeffer2022replication} to capture how stories and memories about benefits and risks of AI spread within the society: (3) Write down a story that you heard from someone about benefits of AI; (4) Write down a story that you heard from someone about risks of AI. To understand participants' views on the \textit{future} of AI systems, we used probing strategies grounded in speculative design~\cite{marenko2018futurecrafting, auger2013speculative, wong2018speculative} and asked: (5) If you had a magic wand that could create an AI system, what would you want that AI system to do for you? (6) How could the AI system that you just described put you (or someone else) at risk? Finally, to explore perceptions of the \textit{trustworthy} development of AI systems, we asked: (7) What characteristics should an AI system have to be trustworthy? For this question, we used a slightly modified version of NIST's definition of AI systems~\cite{standards2023ai}. More details about the survey design are provided below and in Appendix~\ref{app:survey-additional}.

\textbf{Refining the Pilot Survey.}
We employed five strategies to mitigate known issues with survey research in our pilot survey~\cite{gilovich2006being, albert2009beyond, colton2007designing, groves2011survey}: (1) We positioned demographic questions toward the end of our survey to mitigate potential priming and sensitivity concerns that could result from stereotype threat; (2) we incorporated two attention-check questions to identify low-quality responses; (3) we conducted expert reviews with five domain experts before deploying the survey to improve clarity; (4) we did a walk-through with people who have lived most of their lives in countries included in the survey but where authors have not resided, to capture their views on the survey design (e.g., in Australia, people may consider the benefits of AI for various aspects of their lives differently compared to other countries, with a particular emphasis on the importance of indigenous identity recognition in their region; or in Japan, the use of ``stories'' could reflect factual events or rumors/gossips); and (5) we conducted small survey pilots in two rounds (with six and five participants, respectively) to pinpoint areas for further clarification, before conducting the larger pilot.

\textbf{Pilot Deployment.} 
We hosted the survey on Qualtrics~\cite{qualtrics2021} and used Prolific~\cite{prolific2022}, a crowdsourcing platform, for participant recruitment, between July and August~2023. Using Prolific's screening tool and its ``gender-balanced'' sampling,\footnote{Prior work shows a ``gender-balance'' sample on Prolific is similar to its ``representative'' sample but costs less~\cite{tang2022replication}. Therefore, due to budget limits, we opted in for a ``gender-balanced'' sample. We discuss potential harms of this framing in Section~\ref{app:reflections}.} we selected participants who were at least 18 years old and fluent in English, had a minimum approval rate of 95\%, and resided in one of the six countries we recruited our study participants from, including Australia (AU), Chile (CL), Israel (IL), the United Kingdom (UK), the United States (US), and South Africa (ZA). These countries were chosen based on access to participants through our recruitment platform to cover one country per continent.

\textbf{Recruitment.}
We recruited~50 participants from each of the six countries, totaling 300 participants. Six responses were removed due to failed attention checks, and 12 additional responses were discarded as we classified them as either AI-generated or copied from the Internet (based on discussions among authors). The resulting dataset is comprised of 282 responses. Participants were paid \$5 USD via Prolific for completing the study. The survey's average completion time was 22 minutes ($std=11$ minutes), with responses averaging 211 words in length. 

\textbf{Participant Demographics.} 
Our sample returned an almost equal number of participants in each country (see Appendix~\ref{app:demographics} for a summary of participant demographics). We acknowledge that our sample did not attempt to measure \textit{internal} diversity or sample sub-populations within each country, a choice that we discuss in Section~\ref{app:reflections}.

\subsection{Systematic Literature Review}
\label{sec:lit-review}
We conducted a systematic literature review centered on papers related to the themes of \textit{public}, \textit{AI}, \textit{surveys}, and \textit{perceptions} described in a variety of ways (see Appendix~\ref{app:query-review} for search terms). We sourced our material from the ACM DL (conference publications) as well as from Springer's AI \& Ethics and AI \& Society journals (journal publications). Our search was constrained to a two-year period (01/2022--01/2024), except for papers from AIES and FAccT, for which we did not impose any date restrictions. The exclusion criteria were as follows: (1) surveys of literature, policies, or guidelines, rather than respondents; (2) use of past surveys or datasets; (3) surveys not intended to include representative samples or, if purposive, not intended to have large reach; and (4) abstracts or short papers lacking detailed methods (see details in Appendix~\ref{app:prisma}). Our final dataset consists of \textbf{44 papers}, including 15 papers from ACM conferences, 6 from AI \& Ethics, and 23 from AI \& Society. A spreadsheet with a list of all the papers we reviewed, along with our analysis, is available as supplementary material. 

\textbf{Limitations.}
Our keywords span a wide range within each topic of interest, but our search may not encompass all available literature in the domain. Nevertheless, we believe our queries sufficiently capture recent trends in the use of surveys within the AI research community. We acknowledge that our understanding of the literature is influenced by our positionality and academic backgrounds (see Section~\ref{sec:positionality}). Future work could expand on our research to include non-academic literature, such as studies conducted by corporate research platforms like Pew or Gallup, or other agencies worldwide. Our focus was on the recent surge in AI, particularly post-generative AI, with conferences like AIES established to address the ethical implications of AI. Thus, we concentrated on literature from the past two years. Future research could expand this time frame to identify long-term patterns in the use of surveys in computing and AI.

\section{Large-Scale Surveys of AI in the Literature}
\label{sec:lit-review-findings} 
After analyzing our corpus of 44 papers, we noticed inconsistent practices in the reporting of research procedures (e.g., ethical review approvals, informed consents, and concerns related to cultural sensitivity and congruence of research practices) and research transparency practices (e.g., lack of information about funding sources, positionality statements, recruitment strategies, and socio-demographic characteristics of participants or their compensation). 

\textbf{Research Methods.}
Across the 44 papers, there were 58 primarily quantitative studies (some papers reported results of more than one study) conducted (e.g., surveys and online experiments, with qualitative analysis of open-ended responses), nine of which were accompanied by qualitative studies with human participants (e.g., interviews and focus groups, sometimes with quantitative insights into occurrence counting or other descriptive data). 
The median sample size for the quantitative studies was $n=607$ participants ($mean=2,668$, $min=57$, $max=47,951$, $std=8,259$). The median sample size for the qualitative studies was $n=12$ participants ($mean=20$, $min=8$, $max=45$, $std=14$). 

\textbf{Recruitment Strategies.}
Ten out of 44 papers did not report where they recruited their participants from, among those that reported it, the most common recruitment strategy was using a market research agency ($n=11$) and online or crowdsourcing platforms ($n=8$) (e.g., Prolific~\cite{prolific2022}, MTurk~\cite{mturk2024}), or using the census panels ($n=2$) or electoral poll databases ($n=2$). Other recruitment strategies included snowball sampling or word-of-mouth ($n=4$), social media ($n=3$), internal databases, for example, in universities ($n=3$), direct emails ($n=3$), convenience samples ($n=3$), online websites advertising the study ($n=2$), and physical posters ($n=1$). Many of these methods use online channels to recruit participants, which may exclude certain marginalized populations that do not have consistent access to the Internet or lack knowledge, skills, experience, or physical abilities to engage with those online channels. Online panels (especially, MTurk) have been previously criticized for having non-diverse or non-representative user bases~\cite{posch2018characterizing}, for data quality issues, associated with dishonesty, or low-effort responses especially among most experienced survey respondents~\cite{peer2022data,douglas2023data}, and more recently, for using AI for completing tasks~\cite{veselovsky2023artificial}.

\textbf{Reporting of Demographics.}
The reporting of socio-demographic characteristics was not consistent across the papers. Except for gender ($n=36$) and age ($n=34$) that were reported in the majority of papers, and education levels ($n=20$) that were reported in slightly less than half of the papers, other characteristics like race/ethnicity, income, and employment status were reported only in a handful of papers. Fourteen papers examined samples that were census- or nationally-``representative''---one reported using a quota-stratified sample for age and gender, and one recruited participants from the general public but put effort into including marginalized groups such as people of color, gender minorities, and those with mental illnesses. The remaining papers ($n=28$) either used random sampling or did not report sampling strategies; some of these papers reported achieving balanced samples in terms of age and gender, while others had skewed samples. 

\textbf{Geographic Diversity Among Authors.}
Only six papers had authors from the Global South, and 11 papers did not include authors from the countries where these studies were conducted (specifically, two of these 11 studies included populations from the Global South but did not include authors who lived or worked in these countries). These findings suggest trends of limited geographic diversity among authors and raise concerns that authors may not always have the sufficient cultural context about the populations they study. Two of the 44 papers reported authors' Western points of view or limitations of their stance as computing educators. There is a chance that the authors were born or have lived in countries other than the countries of authors' current affiliations and, therefore, have sufficient cultural context. However, without a positionality statement, it is hard to understand if this is actually the case. Therefore, these findings highlight both the need to strongly encourage positionality statement disclosures when studying human participants and the importance of further in-depth research to address the apparent misalignment between the geographic distribution of authors and the populations they study.

\textbf{Funding Sources.}
Five papers had the opposite issue; these studies had authors from the countries that did not include participants from those countries (e.g., authors from Japan ran a study with US participants, but not with participants from Japan). Many papers acknowledged funding support either from non-profit organizations ($n=14$) like foundations, philanthropic funds, trusts, which are mostly funded by contributions from individual donors and organizations, or from university-sponsored ($n=5$) and government-provided grants ($n=13$), a significant portion of which comes from tax payers' money and other public sources. Five papers acknowledged support from private companies, raising concerns about potential conflicts of interest (as already surfaced by \citet{birhane2022values}, and issues extensively discussed in \citet{young2022confronting}). Finally, 15 papers did not disclose their funding sources.

\textbf{Continuity of Research.}
Most studies in our corpus were cross-sectional (i.e., the data were collected from participants at a single point in time, without periodically repeated measurements). Only one paper reported several (four) rounds of a survey; another reported findings before and after the 2020 US general elections. While one other paper ran a survey similar to another earlier one conducted by the same authors, it did not report any direct comparisons. Two other papers mentioned that their questions were part of an annual survey but did not specify if the same questions were asked again or whether they observed any shift in responses over time, and no precedent or follow-up papers for those annual surveys came up in our literature search. These findings suggest a lack of continuity of existing survey research in AI and a lack of comparability in both repeated surveys and related surveys in the literature.

\textbf{Research Ethics.}
Many papers ($n=20$) did not mention Institutional Review Board (IRB) or other ethical committee approvals, nor did they indicate that informed consent was obtained from participants before the studies, echoing recent findings in AI research with human participants~\cite{mckee2023human}. However, these results may require a more critical view. Researchers report diversity in IRB practices around the world~\cite{patel2013variations, pe2019inconsistencies}. While IRBs are common in the US, there may be different structures in Europe, Asia, and Africa~\cite[e.g.,][]{orimadegun2020protocol}. The practicalities of review and consent may depend on prior experiences of marginalization and exploitation~\cite{angal2016ethics, kuhn2020indigenous, norton1996research, tapaha2017we}. \citet{schrag2010ethical} summarized some of these risks in the term ``\textit{ethical imperialism}''; i.e., the imposition of review board practices from the Global North—particularly from the US—on other countries and cultures, despite differences in local values and practices.

\textbf{Participant Compensation.}
The majority of papers ($n=35$) did not report if participants were compensated for their time. Two papers explicitly stated that participants were not compensated. Four papers reported that participants were compensated, but either did not report how much they were paid or did not explain if the payment was at or above the regional minimum wage levels. Lack of information about compensation obstructs the understanding of incentive mechanisms, and how those could have affected the results. For example, prior research has shown that low compensation (in comparison to the time required to complete the survey) is perceived as unfair by respondents and results in reduced data quality~\cite{lovett2018data}.

\textbf{Testing and Feedback.}
Most of the papers ($n=38$) did not report any testing of the study materials before data collection. Only a few papers mentioned that pilot surveys ($n=4$) or walk-through interviews/focus groups ($n=6$) were conducted prior to launching the main studies. In addition, one paper mentioned expert consultation to review the study materials, and another paper mentioned pre-registering the study on the Open Science Framework's website in order to increase research transparency by specifying analysis methods before analyzing the data. We recognize, however, that this emphasis on pre-testing and pre-registration may inadvertently suggest a rigid, positivist approach to research, potentially overlooking the value of exploratory studies and the interpretive role of researchers. It is therefore important to state that while these practices contribute to the rigor and transparency of research, they are not the sole arbiters of validity, and a balance should be struck to allow for a diversity of methodological approaches.


\textbf{Evaluation and Replication.}
Some papers included study materials such as survey instruments and interview guides ($n=9$)
, full replication packages ($n=2$), or study data ($n=4$), or made the study data available upon request ($n=9$). However, 16 papers did not include additional artifacts, contributing to ongoing concerns about the replication challenge, e.g., lack of replication packages, analysis codes, or datasets~\cite{dreber2019statistical,freese2017replication,echtler2018open}.

\section{Discussion}
\label{sec:why-harmful}

\subsection{Pitfalls of Surveys as Enablers of ``Participation''} 

A recent trend in critical AI research advocates for \textit{participation} in AI~\cite[e.g.,][]{feffer2023from, sloan2022participation, delgado2023participatory, bondi2021envisioning}, with researchers using different terminologies and viewpoints to describe levels of participation, with frequent reference to the ladder of participation~\cite{arnstein1969ladder}. For example, \citet{sloan2022participation} categorize participation in AI into three levels: \textit{work}, \textit{consultation}, and \textit{justice}, whereas \citet{delgado2023participatory} propose a four-level framework: \textit{consult}, \textit{include}, \textit{collaborate}, and \textit{own}. Given that surveys are often a paid, one-time transaction, they most likely fall under the categories of participation as \textit{work} and \textit{consultation}---the minimum level of participation in both frameworks. Other frameworks for assessing participation in AI propose questions for researchers to consider. \citet{birhane2022power} provide a framework focusing on \textit{empowerment}, \textit{reciprocity}, and \textit{reflexivity} for critically evaluating participatory approaches in AI. \citet{feffer2023from} offer a more detailed ten-axis framework that encompasses \citeauthor{birhane2022power}'s dimensions as well. We adopt the ten-axis framework by \citet{feffer2023from} to assess surveys as a method for participation in AI given its comprehensive integration into prior scholarship:

\textbf{1. Representation.} As a term, representation has often been misused in public survey literature regarding opinions on AI, which can create an \textit{illusion} of rigor, act as a veneer for bias reduction, and give misleading perceptions of achievements in sampling. In our pilot survey, we could have used terms like ``cross-cultural'' or ``across six continents'' in our title or abstract to attract reader's attention, while the results did not truly speak to these terms.
    
\textbf{2. Stage.} The current literature primarily focuses on early stages of design, development, and governance of AI to capture attitudes and perceptions, to inform either products or policies. Surveys, like ours, often focus on capturing what people think about AI technologies, such as autonomous driving.~\cite{etienne2024autonomous, awad2020crowdsourcing}, which is intended to inform product design or policies, or to help with ``\textit{identifying citizens' expectations.}''~\cite{awad2020crowdsourcing}

\textbf{3. Setting.} In reviewed papers, the research setting was situated online, with unknown, paid, or free compensation structures. We had to decide how to fairly compensate participants across countries (e.g., \$10 per hour does not provide the same level of compensation in the US and Chile, due to differences in living costs and minimum wages).

\textbf{4. Resources.} While prior research has often failed to consistently report the details, in our survey, the primary resources needed to conduct high-quality research were detailed, including the reliable channels used to recruit participants, digital consent and survey instruments, and the budget to cover research expenses (such as participant compensation, platform fees, researcher salaries, and costs of analysis software licenses). Additionally, hosting shareable versions of survey results for respondents and other communities also necessitates resources, as outlined below.

\textbf{5. Communication.} Communication between researchers and participants is often, unidirectional, with a transactional or task-oriented tone. There is no way for participants to know about the results without actively seeking and searching for results of their participation. In Prolific, the platform we used for our pilot survey, there is a feature that allows for messaging between participants and researchers. However, it is not intended for discussing the results of the survey; rather, it is designed for addressing any issues related to responses or payments. Recent work by \citet{do2024gigsousveillance} suggests that Prolific is like other gig work platforms, which similarly limit communications among workers and employers to transactional matters, and which discourage or prohibit communications from one gig worker to another.

\textbf{6. Elicitation.} Surveys are generally not structured to be longitudinal and are seldom designed to engage participants through multiple methods. In online surveys, participants are often given a limited amount of time to complete the survey. In particular, if they are professional participants, they may take multiple surveys in a short period of time~\cite{hara2018data}. This scenario can lead to fatigue and diminish the level of meaningful engagement with the survey.

\textbf{7. Conflict resolution.}
Survey data is usually aggregated and converted into numbers and statistics, representing a form of knowledge. Surveys are not typically seen as collaborative methods that involve ongoing discussions or conflict resolution processes. Consequently, they do not incorporate participant feedback or clarifications after data collection, nor do they include participants' input on the researchers' analytical summaries of results. Although our pilot survey featured multiple open-ended questions requiring thematic analysis, which necessitated reflection and collaboration among researchers, this approach still did not allow participants to contribute to the analysis.

\textbf{8. Feedback.}
Although surveys may include a final question asking for participants' overall feedback or comments about the study, or provide an option to message or email the researcher, there is typically no mechanism for participants to access the results, offer feedback, or influence the final report or its implications. In our consent form, we provided an email address for participants to ask questions; however, no one reached out to us regarding the study.

\textbf{9. Empowerment.}
The potential outcomes for participants and how the results might benefit or pose risks to them are often unclear. This information could be conveyed in the consent form, but many papers we reviewed lacked these details. In our pilot study, which focused mainly on understanding people's perceptions, there is no apparent direct benefit to the participants. While researchers may gain from publications that cover large segments of certain populations, and companies may benefit from extensive datasets reflecting attitudes toward their AI technologies, it remains unclear how participants are empowered through the collection and use of survey data.

\textbf{10. Evaluation.}
Transparency regarding the details of methods and analyses, or the availability of full replication packages, is necessary to evaluate research practices. However, these elements were often missing from the papers we reviewed. Additionally, making data publicly available poses challenges from a researcher’s perspective, as IRBs are frequently cautious about data-sharing practices, particularly concerning qualitative data. Open data are vulnerable to privacy violations~\cite{borgesius2015open, crawford2014big}---particularly in medical domains---and to cultural misappropriation, extraction, and exploitation \cite{duarte2021native, karsgaard2023new, palacios2022note}. Our pilot survey's consent form and IRB application indicated that anonymized data would be made public to ensure transparency and facilitate future evaluation. We also planned to include the details of our qualitative coding in the replication package.

Current research practices and platforms have a significant impact on how surveys are assessed. To maintain anonymity, researchers and participants are often completely disconnected, leaving participants with no way to receive feedback on their contributions or learn about the results. Even if there were a way for participants to access the outcomes, traditional research publications would not be the most effective means of dissemination. These publications are often lengthy, use scientific jargon, require a high reading level, and are frequently behind paywalls. Additionally, the AI research community often emphasizes technological impacts—such as creating new datasets or enhancing efficiency—over the societal impacts of their research~\cite{birhane2022values}. Therefore, establishing a bidirectional communication and feedback loop in survey research necessitates a fundamental shift in how research impact is perceived and evaluated within the community.

Researchers have the discretion to use, report, or disregard parts of the results. What happens if researchers disagree with participants' responses, or if the outcomes do not align with the research goals or hypotheses? How much influence do participants have over the results and their eventual impact? Under current research practices, especially in online settings, participants have no power to affect or control the implications of their contributions. Such large-scale surveys are employed to \textit{inform} rather than \textit{empower}, treating survey takers as ``\textit{subjects in an investigation}''~\cite{himmelreich2023against}.

\subsection{Representation Issues in Surveys}
\label{sec:ref-representation-harmful}

\subsubsection{True Representation or an Illusion of Representation?}
Dataset attributes and annotation practices are known to introduce biases into AI, potentially resulting in poor representation of the perspectives of marginalized groups~\cite{bergman2023representation, d_ignazio2020data, sambasivan2021everyone, lu2024perceptions}. A model trained on a ``representative'' dataset might still under-perform when making decisions for marginalized communities~\cite{aragon2022human, bergman2023representation}. Our pilot survey, which involved 282 participants, was methodologically acceptable, as we continued recruiting until data saturation---an acceptable practice in qualitative research~\cite{francis2010what, guest2006how}. However, it does not fully encompass the entire spectrum of people in the studied countries, which are likely to be internally heterogeneous, potentially including marginalized or minoritized groups---a situation described by \citet[p. 1]{trefzer2014introduction} as ``\textit{the global south within the global north.}'' In our literature review, 14 out of 44 papers (32\%) claimed to have a ``nationally-representative'' sample. The remaining papers typically included over 300 participants (with a median of 607 participants), aligning with the recommended sample size of 385 participants for a large population with 95\% confidence and a 5\% margin of error~\cite{surveymonkey2024calculate}.

While survey methods can be effective for specific studies with targeted design and population, such as testing a new app feature, they may not be adequate for complex and impactful topics like AI, where results could affect large populations. Our work extends the discussion of representation issues to survey methods. We critique how surveys influence our research, development, and governance of AI, highlighting issues that extend beyond datasets and model development to encompass how AI should \textit{be} and \textit{behave}. Companies like Anthropic base their AI models' behavior on survey responses~\cite{anthropic2023}, and Google glorifies public optimism about AI using their large-scale survey in partnership with Ipsos~\cite{google2024global}. Conversely, other surveys by Ipsos show increasing public nervousness about AI~\cite{boyon2023nervous}, with opinions varying depending on a country's level of economic development~\cite{boyon2022opinions}. In this line of research, \citet{feffer2023from} critique the Moral Machine experiment, which claims to represent global perspectives on the ethics of machines by encompassing 233 countries and 40 million decisions. However, an examination of the survey's geographical coverage reveals that it is largely dominated by contributions from North America, Europe, and some parts of South America, with minimal input from the African continent. These discrepancies in ``representative'' surveys and claims regarding ``representative'' results reinforce our argument about the true meaning of representation and underscore the need for our community to critically assess the current use of survey methods.

\subsubsection{Cross-Cultural Surveys Considered Helpful or Harmful?}
The complexity increases when conducting surveys across different cultures and regions \cite{duarte2021native, karsgaard2023new, palacios2022note, davani2024disentangling}. In our pilot, we aimed to capture a broad range of perspectives from six continents without being fully immersed in the target populations and experiencing what it means to live in countries like Chile or South Africa. Or, conducting surveys exclusively in English in countries, where this is not the official language, limited our insights and was exclusionary by design. We received responses in languages other than English, indicating a desire to participate irrespective of English proficiency, driven by interest or compensation. Similarly, 11 out of 44 papers (25\%) in our dataset studied countries without having an author affiliated with those countries. 

The literature review also validates prior concerns about research focusing predominantly on Western populations~\cite{septiandri2023weird}---only six papers (14\%) featured authors from the Global South. This lack of representation could create an echo chamber effect, where the voices of certain populations dominate the discourse in AI (including responsible AI, safety of AI, and ethics of using AI), potentially eclipsing perspectives from underrepresented regions or populations \cite{duarte2021native, trefzer2014introduction}. While the intention to include more countries might be well-meaning, the way \textit{how} this inclusivity is approached remains a topic that requires attention and care. Without meaningfully engaging with the target populations, conducting cross-cultural studies might be motivated by the desire for a large sample size rather than an understanding of cross-cultural differences, potentially causing harm rather than benefit due to misinterpretations and a lack of contextual understanding, including different conceptions of the meaning of \textit{consent} based on cultural roles of decision-makers and/or histories of exploitation \cite{munteanu2019field}.

Another concern is the use of standardized questions across different cultures. Employing a uniform set of questions, even with thorough translations, suggests that we, as researchers, have not adequately adapted our inquiries to specific cultures and communities. This approach risks missing key cultural insights and misinterpreting responses rooted in specific cultural contexts. While standard questions aid comparative analysis, the extent to which they overlook contextual nuances and local values is an open question. This highlights the need not only for localization of study materials but also for more flexible and culturally-sensitive research methodologies in general, especially when studying diverse populations. When we sought feedback on our pilot survey, we found that people had diverse interpretations of the questions. For example, the term \textit{story} can have various meanings in Japanese, and a specific demographic question about ethnic origins might be crucial for indigenous populations in Australia. However, as researchers without lived experiences in the target countries, we would not have been able to understand these cultural details without conducting walk-through interviews with representatives from those countries—a practice that was absent in 85\% of the papers we reviewed.

\subsection{Value Tensions in Surveys: Heuristic Questions}
\label{sec:value-tensions}

In this paper, we discussed many positions and concerns, which often did not converge toward simple outcomes. Survey research in AI vexes us with many choices and decisions. We summarize the major issues as tensions that are currently unresolved. In view of the many reasons and motivations for using surveys in AI research, we propose that ``\textit{heuristic questions}''~\cite{muller1997ethnocritical} may be more valuable than advice. Asking ``\textit{big questions}''~\cite[ms. p. 1]{beck2017reviewing}, \cite{reiser2017asking, schaeffer2003science} or ``\textit{the right question}''~\cite[ms. p. 1]{mao2019data} has been deemed valuable when facing new or newly-problematic research challenges~\cite{phillips2018beyond}. Thus, we propose to use the following reflexive questions~\cite[e.g., inspired by][]{laufer2022four, septiandri2023weird} when planning survey-based studies about AI:

    \textbf{(A) Breadth and Depth:} 
    \begin{itemize}
        \item \textbf{Standardization and Customization.} Do we attempt to standardize certain survey content through invariant questions addressed to all persons in all geographical locations and cultural backgrounds~\cite{ornstein2013companion}? How can we address the tendency for a survey to primarily reflect the cultural perspective of the Global North and its associated dominance~\cite{septiandri2023weird}?
        
        \item \textbf{Languages.} Do we adapt the survey for regional languages~\cite{kelley2021exciting}? When does openness and accommodation shade into cultural differences in inquiry, leading to incommensurable outcomes?
    
        \item \textbf{Sampling.} Survey research is often constrained by time and budget. In settings with diverse cultures, how great an effort should be spent on recruiting a balanced or weighted sample across cultures~\cite{selwyn2020ai, zhang2020us}? Where is the ``\textit{stopping point}'', and is this a question that requires members of the survey population to help answer? How much sampling stratification is needed? Who defines cultural boundaries?
    \end{itemize}

    \textbf{(B) Manual and Automated Approaches.} If we use generative AI in question generation, do we risk a bland, so-called ``\textit{universal}'', tone that reflects the worldview of the AI-provider~\cite{paxton2023like}? Or if we rely on humans in the research process, e.g., in data collection and analysis, how can we account for the limited views of research teams and their biases? What are the specific choices related to use of AI that need to be disclosed as part of consent~\cite[e.g.,][]{wilcox2023aiconsent, andreotta2022ai, gomez2023beyond} and transparency in publication~\cite[e.g.,][]{wacharamanotham2020transparency, hosseini2023ethics}?
    
    \textbf{(C) Who and What Influences Survey Designs?} Survey design is often considered to be the domain of specialists~\cite[e.g.,][]{fink2003design, spector2013survey}. While it is true that the design of questions and response-scales requires professional knowledge, the selection of topics may be informed by members of stakeholder groups or affected classes~\cite{baeza_yates2018bias, bloomberg2023humans, garcia2021decolonial, bird2020decolonising, kwet2019digial}. Mindful of the different meanings of ``\textit{participation}''~\cite[e.g.,][]{hansen2021third, muller1993participatory, schuler1993participatory, simonsen2012routledge}, we ask: What are the opportunities for involving community members or leaders in participatory or co-design processes to select and refine the survey topics and the way they are framed in questions~\cite[e.g.,][]{arnstein1969ladder, dugan2021participatory, flicker2010survey,schulz2005cbpr}?
    
    \textbf{(D) Trust and Research Engagement.} Trust among many communities that might participate in survey research has been eroded due to past mistreatment in research more broadly, along with experiences of racism and various forms of prejudice by different research institutions~\cite{scharff2010more}. This historical context influences a community's decisions regarding whether and how to engage in such activities.~\cite{wilcox2023infra}. To what extent should survey design, deployment, and reporting directly address issues of trust? What measures (e.g., establishing publication standards that include sharing data with participants) should we expect our community to implement to ensure responsible and transparent research practices?
    
    \textbf{(E) Mixed Methods and Balanced Inquiry.} As discussed in Section~\ref{sec:introduction}, surveys tend to decontextualize responses and isolate respondents~\cite{ornstein2013companion}. Is it feasible to integrate large-scale quantitative survey methods with smaller-scale, rigorous qualitative analyses involving strategically selected groups of informants~\cite[e.g.,][]{baumer2017comparing, greenberg2008usability, muller2016machine}?

    \textbf{(F) Transparency and Research Practices.} We should discuss whether and how to establish consistent reporting methods for surveys on AI topics. For instance, what transparency artifacts~\cite[e.g.,][]{crisan2022interactive, mitchell2019model, chmielinski2022dataset,CrowdSheets, Healthsheet, Artsheets} might inspire new forms of methodological transparency? How should we determine which types of data to include in these artifacts? Selection criteria to consider may include the contingent and potentially sensitive nature of AI topics addressed in surveys, the diverse communities that survey research serves, the fact that cross-cultural perspectives may require varied forms of transparency, the need for participant anonymization, and the interpretive traditions associated with qualitative analyses~\cite{soden2024evaluating}.

    \textbf{(G) Researcher and Participant Empowerment.}
    Researchers should consider early in their study what survey participants stand to gain from the research~\cite{oldendick2012survey, jamieson2023reflexivity}. This consideration becomes particularly important when studying hard-to-reach populations or when using public funds (or considering conflicts of interest when using private funds). Adopting a reflexive, critical approach to the implications of their research can significantly benefit researchers in understanding and improving the value of their work for participants. In reflecting, we ask: Are there alternative avenues to improve the value exchange for the populations being studied?

\section{Conclusion}
In this paper, we combine epistemic approaches grounded in critical reflexivity with a systemic literature review to examine the state of large-scale surveys in AI scholarship. The study reveals a range of performative and misleading practices with a method that has garnered adoption, informing research, shaping publicly-facing narratives, and justifying trajectories in AI development---highlighting the need for urgent intervention. The stakes are high in AI research, and some of these issues cannot be adequately addressed or dismissed by merely tucking challenges within the limitations section of research papers. As such, we aim to spark reflexive engagement with research processes that shape how surveys could be used responsibly and offer a list of heuristic questions to prompt more thorough acknowledgments of bias and subjectivity.


\section{Research Ethics and Social Impact}

\subsection{Ethical Considerations Statement}
In addition to the ethics considerations described in our paper body, our pilot survey obtained approval from the Research Ethics Office at King's College London. We implemented strict measures to ensure the confidentiality, anonymity, and privacy of our participants. No personally-identifiable information was collected, and participation was voluntary and anonymous. Participants were provided with an informed consent form in English, detailing the study's purpose and the intended use of the data collected.

\subsection{Researcher Positionality Statement}
\label{sec:positionality}
The authors come from varied research backgrounds that shape their perspectives. The study was funded by an academic institution in the Global North, and funding for the study was restricted to respondent incentives and vendor survey services. Authors were employed by their institutions and were not explicitly paid to conduct this research. One author is employed by an academic institution, and one is employed at a non-governmental organization. Four authors are employed in industry research roles, though this study was not part of their company research. The research team have experiences living in two of the six countries surveyed (United Kingdom and United States). All six authors have extensive experience with survey methods, with four having experience with international and cross-cultural survey approaches. Authors were born in, currently live in, or had previously lived in, nine different countries collectively. Our race/ethnicity is collectively White (European) ($n=3$), Middle Eastern ($n=2$), and Afro-Caribbean ($n=1$). All authors identify as having some experience with marginalization in computing, either through years of conducting computing research with marginalized groups or as members of a marginalized group themselves.

Our positionality is influenced by our backgrounds and experiences; as researchers trained and working in predominantly Western institutions, we acknowledge that complementary scholarship related to our research questions is needed, to further the understandings presented in this paper. Our positionality has also influenced the subjectivity inherent in framing our paper approach, research questions, study pilot design, literature review, and data interpretation and analysis, as we elaborate on throughout the paper.

\subsection{Adverse Impact Statement}
Our research aims to promote critical thinking within the AIES community about survey methods in AI, but they could be interpreted as an outright dismissal of these methods without full engagement with the nuances we present in our paper. Our intention is not to entirely discourage the use of surveys in AI and responsible AI research. Instead, our goal is to foster thoughtful and critical engagement within the AIES community to develop perspectives on the principles associated with the \textit{who, what, when, where, why,} and \textit{how} of human survey methods in AI research. Finally, we do not intend to suggest that designing the ``\textit{perfect}'' survey will address the systemic issues that surround survey research in AI. Power relationships and broader structural concerns cannot be resolved simply by a survey designed and deployed in ways that uphold our community's principles; though such a survey may meet specific research goals, it would not address issues surrounding its application (e.g., the actual impact that survey results have on the practices of powerful institutions). We encourage future research to continue utilizing reflexive approaches and to further develop best practices and standards for conducting high-quality, inclusive, reliable, and impactful survey research in AI.

\bibliography{biblio}

\begin{thebibliography}{238}
\providecommand{\natexlab}[1]{#1}

\bibitem[{{Ada Lovelace Institute}(2021)}]{ADA-Participatory-Data}
{Ada Lovelace Institute}. 2021.
\newblock Participatory data stewardship: A framework for involving people in the use of data.

\bibitem[{{Ada Lovelace Institute and Alan Turing Institute}(2022)}]{ada2023UK}
{Ada Lovelace Institute and Alan Turing Institute}. 2022.
\newblock How do people feel about {AI}?

\bibitem[{Agyemang et~al.(2009)Agyemang, Addo, Bhopal, de~Graft~Aikins, and Stronks}]{agyemang2009cardiovascular}
Agyemang, C.; Addo, J.; Bhopal, R.; de~Graft~Aikins, A.; and Stronks, K. 2009.
\newblock Cardiovascular disease, diabetes and established risk factors among populations of sub-Saharan African descent in Europe: a literature review.
\newblock \emph{Globalization and health}.

\bibitem[{AI.gov(2023{\natexlab{a}})}]{govai2023NAIAC}
AI.gov. 2023{\natexlab{a}}.
\newblock National Artificial Intelligence Advisory Committee (NAIAC).

\bibitem[{AI.gov(2023{\natexlab{b}})}]{govai2023strengthening}
AI.gov. 2023{\natexlab{b}}.
\newblock Strengthening and Democratizing the U.S. Artificial Intelligence Innovation Ecosystem --- An Implementation Plan for a National Artificial Intelligence Research Resource.

\bibitem[{Albert, Tullis, and Tedesco(2009)}]{albert2009beyond}
Albert, B.; Tullis, T.; and Tedesco, D. 2009.
\newblock \emph{Beyond the usability lab: Conducting large-scale online user experience studies}.
\newblock Morgan Kaufmann.

\bibitem[{Alexander and Moore(2021)}]{sep-ethics-deontological}
Alexander, L.; and Moore, M. 2021.
\newblock Deontological Ethics.
\newblock In Zalta, E.~N., ed., \emph{The {Stanford} Encyclopedia of Philosophy}. Metaphysics Research Lab, Stanford University, {W}inter 2021 edition.

\bibitem[{Ali(2016)}]{ali2016does}
Ali, S.~M. 2016.
\newblock A Brief Introduction to Decolonial Computing.
\newblock \emph{XRDS}.

\bibitem[{Alkhathlan et~al.(2024)Alkhathlan, Cachel, Shrestha, Harrison, and Rundensteiner}]{alkhathlan2024balancing}
Alkhathlan, M.; Cachel, K.; Shrestha, H.; Harrison, L.; and Rundensteiner, E. 2024.
\newblock Balancing Act: Evaluating People’s Perceptions of Fair Ranking Metrics.
\newblock In \emph{The 2024 ACM Conference on Fairness, Accountability, and Transparency}.

\bibitem[{Alvarado~Garcia et~al.(2021)Alvarado~Garcia, Maestre, Barcham, Iriarte, Wong-Villacres, Lemus, Dudani, Reynolds-Cu\'{e}llar, Wang, and Cerratto~Pargman}]{garcia2021decolonial}
Alvarado~Garcia, A.; Maestre, J.~F.; Barcham, M.; Iriarte, M.; Wong-Villacres, M.; Lemus, O.~A.; Dudani, P.; Reynolds-Cu\'{e}llar, P.; Wang, R.; and Cerratto~Pargman, T. 2021.
\newblock Decolonial Pathways: Our Manifesto for a Decolonizing Agenda in HCI Research and Design.
\newblock In \emph{Extended Abstracts of the 2021 CHI Conference on Human Factors in Computing Systems}. ACM.

\bibitem[{Amazon(2023)}]{mturk2024}
Amazon. 2023.
\newblock Amazon Mechanical Turk.

\bibitem[{Andreotta, Kirkham, and Rizzi(2022)}]{andreotta2022ai}
Andreotta, A.~J.; Kirkham, N.; and Rizzi, M. 2022.
\newblock AI, big data, and the future of consent.
\newblock \emph{Ai \& Society}.

\bibitem[{Angal et~al.(2016)Angal, Petersen, Tobacco, Elliott, in~SIDS, and Network}]{angal2016ethics}
Angal, J.; Petersen, J.~M.; Tobacco, D.; Elliott, A.~J.; in~SIDS, P.~A.; and Network, S. 2016.
\newblock Ethics review for a multi-site project involving Tribal Nations in the Northern Plains.
\newblock \emph{Journal of Empirical Research on Human Research Ethics}.

\bibitem[{Anthropic(2023)}]{anthropic2023}
Anthropic. 2023.
\newblock Collective Constitutional AI: Aligning a Language Model with Public Input.

\bibitem[{Aragon et~al.(2022)Aragon, Guha, Kogan, Muller, and Neff}]{aragon2022human}
Aragon, C.; Guha, S.; Kogan, M.; Muller, M.; and Neff, G. 2022.
\newblock \emph{Human-centered data science: An introduction}.
\newblock MIT Press.

\bibitem[{Arai and Matsumoto(2023)}]{arai2023public}
Arai, K.; and Matsumoto, M. 2023.
\newblock Public perceptions of autonomous lethal weapons systems.
\newblock \emph{AI and Ethics}.

\bibitem[{Arcaya et~al.(2018)Arcaya, Schnake-Mahl, Binet, Simpson, Church, Gavin, Coleman, Levine, Nielsen, Carroll et~al.}]{arcaya2018community}
Arcaya, M.~C.; Schnake-Mahl, A.; Binet, A.; Simpson, S.; Church, M.~S.; Gavin, V.; Coleman, B.; Levine, S.; Nielsen, A.; Carroll, L.; et~al. 2018.
\newblock Community change and resident needs: designing a participatory action research study in metropolitan Boston.
\newblock \emph{Health \& place}.

\bibitem[{Arnstein(1969)}]{arnstein1969ladder}
Arnstein, S.~R. 1969.
\newblock A ladder of citizen participation.
\newblock \emph{Journal of the American Institute of planners}.

\bibitem[{Auger(2013)}]{auger2013speculative}
Auger, J. 2013.
\newblock Speculative design: crafting the speculation.
\newblock \emph{Digital Creativity}.

\bibitem[{Awad et~al.(2020)Awad, Dsouza, Bonnefon, Shariff, and Rahwan}]{awad2020crowdsourcing}
Awad, E.; Dsouza, S.; Bonnefon, J.-F.; Shariff, A.; and Rahwan, I. 2020.
\newblock Crowdsourcing moral machines.
\newblock \emph{Commun. ACM}.

\bibitem[{Awad et~al.(2018)Awad, Dsouza, Kim, Schulz, Henrich, Shariff, Bonnefon, and Rahwan}]{awad2018moralmachine}
Awad, E.; Dsouza, S.; Kim, R.; Schulz, J.; Henrich, J.; Shariff, A.; Bonnefon, J.-F.; and Rahwan, I. 2018.
\newblock The Moral Machine experiment.
\newblock \emph{Nature}.

\bibitem[{Baeza-Yates(2018)}]{baeza_yates2018bias}
Baeza-Yates, R. 2018.
\newblock Bias on the Web.
\newblock \emph{Commun. ACM}.

\bibitem[{Bao et~al.(2022)Bao, Krause, Calice, Scheufele, Wirz, Brossard, Newman, and Xenos}]{bao2022whose}
Bao, L.; Krause, N.~M.; Calice, M.~N.; Scheufele, D.~A.; Wirz, C.~D.; Brossard, D.; Newman, T.~P.; and Xenos, M.~A. 2022.
\newblock Whose AI? How different publics think about AI and its social impacts.
\newblock \emph{Computers in Human Behavior}.

\bibitem[{Bartneck, Yogeeswaran, and Sibley(2023)}]{bartneck2023personality}
Bartneck, C.; Yogeeswaran, K.; and Sibley, C.~G. 2023.
\newblock Personality and demographic correlates of support for regulating artificial intelligence.
\newblock \emph{AI and Ethics}.

\bibitem[{Baumer et~al.(2017)Baumer, Mimno, Guha, Quan, and Gay}]{baumer2017comparing}
Baumer, E.~P.; Mimno, D.; Guha, S.; Quan, E.; and Gay, G.~K. 2017.
\newblock Comparing grounded theory and topic modeling: Extreme divergence or unlikely convergence?
\newblock \emph{Journal of the Association for Information Science and Technology}.

\bibitem[{Beck and Stolterman(2017)}]{beck2017reviewing}
Beck, J.; and Stolterman, E. 2017.
\newblock Reviewing the big questions literature; or, should HCI have big questions?
\newblock In \emph{Proceedings of the 2017 Conference on Designing Interactive Systems}.

\bibitem[{Beck(1992)}]{beck1992risk}
Beck, U. 1992.
\newblock \emph{Risk society: Towards a new modernity}.
\newblock sage.

\bibitem[{Beck(1996)}]{beck1996world}
Beck, U. 1996.
\newblock World risk society as cosmopolitan society? Ecological questions in a framework of manufactured uncertainties.
\newblock \emph{Theory, culture \& society}.

\bibitem[{Bergman et~al.(2023)Bergman, Hendricks, Rauh, Wu, Agnew, Kunesch, Duan, Gabriel, and Isaac}]{bergman2023representation}
Bergman, A.~S.; Hendricks, L.~A.; Rauh, M.; Wu, B.; Agnew, W.; Kunesch, M.; Duan, I.; Gabriel, I.; and Isaac, W. 2023.
\newblock Representation in AI Evaluations.
\newblock In \emph{Proceedings of the 2023 ACM Conference on Fairness, Accountability, and Transparency}. ACM.

\bibitem[{Bernstein and Bernstein(1996)}]{bernstein1996against}
Bernstein, P.~L.; and Bernstein, P.~L. 1996.
\newblock \emph{Against the gods: The remarkable story of risk}.
\newblock Wiley New York.

\bibitem[{Berry-James, Gooden, and Johnson~III(2020)}]{berry2020civil}
Berry-James, R.~M.; Gooden, S.~T.; and Johnson~III, R.~G. 2020.
\newblock Civil Rights, Social Equity, and Census 2020.
\newblock \emph{Public Administration Review}.

\bibitem[{Bharadiya(2023)}]{bharadiya2023artificial}
Bharadiya, J. 2023.
\newblock Artificial Intelligence in Transportation Systems A Critical Review.
\newblock \emph{American Journal of Computing and Engineering}.

\bibitem[{Bhopal et~al.(2004)Bhopal, Vettini, Hunt, Wiebe, Hanna, and Amos}]{bhopal2004review}
Bhopal, R.; Vettini, A.; Hunt, S.; Wiebe, S.; Hanna, L.; and Amos, A. 2004.
\newblock Review of prevalence data in, and evaluation of methods for cross cultural adaptation of, {UK} surveys on tobacco and alcohol in ethnic minority groups.
\newblock \emph{BMJ}.

\bibitem[{Bird(2020)}]{bird2020decolonising}
Bird, S. 2020.
\newblock Decolonising Speech and Language Technology.
\newblock In Scott, D.; Bel, N.; and Zong, C., eds., \emph{Proceedings of the 28th International Conference on Computational Linguistics}. International Committee on Computational Linguistics.

\bibitem[{Birhane et~al.(2022{\natexlab{a}})Birhane, Isaac, Prabhakaran, Diaz, Elish, Gabriel, and Mohamed}]{birhane2022power}
Birhane, A.; Isaac, W.; Prabhakaran, V.; Diaz, M.; Elish, M.~C.; Gabriel, I.; and Mohamed, S. 2022{\natexlab{a}}.
\newblock Power to the People? Opportunities and Challenges for Participatory AI.
\newblock In \emph{Equity and Access in Algorithms, Mechanisms, and Optimization}. ACM.

\bibitem[{Birhane et~al.(2022{\natexlab{b}})Birhane, Kalluri, Card, Agnew, Dotan, and Bao}]{birhane2022values}
Birhane, A.; Kalluri, P.; Card, D.; Agnew, W.; Dotan, R.; and Bao, M. 2022{\natexlab{b}}.
\newblock The Values Encoded in Machine Learning Research.
\newblock In \emph{2022 {ACM} {Conference} on {Fairness}, {Accountability}, and {Transparency}}. ACM.

\bibitem[{Blodgett and Madaio(2021)}]{blodgett2021risks}
Blodgett, S.~L.; and Madaio, M. 2021.
\newblock Risks of AI Foundation Models in Education.

\bibitem[{Boholm(1996)}]{boholm1996risk}
Boholm, {\AA}. 1996.
\newblock Risk perception and social anthropology: Critique of cultural theory.
\newblock \emph{Ethnos}.

\bibitem[{Bondi et~al.(2021)Bondi, Xu, Acosta-Navas, and Killian}]{bondi2021envisioning}
Bondi, E.; Xu, L.; Acosta-Navas, D.; and Killian, J.~A. 2021.
\newblock Envisioning Communities: A Participatory Approach Towards AI for Social Good.
\newblock In \emph{Proceedings of the 2021 AAAI/ACM Conference on AI, Ethics, and Society}. Association for Computing Machinery.

\bibitem[{Borgesius, Gray, and Van~Eechoud(2015)}]{borgesius2015open}
Borgesius, F.~Z.; Gray, J.; and Van~Eechoud, M. 2015.
\newblock Open data, privacy, and fair information principles: Towards a balancing framework.
\newblock \emph{Berkeley Technology Law Journal}.

\bibitem[{Bourdieu(1990)}]{bourdieu1990other}
Bourdieu, P. 1990.
\newblock \emph{In other words: Essays toward a reflexive sociology}.
\newblock Stanford University Press.

\bibitem[{Bourdieu and Wacquant(1992)}]{bourdieu1992invitation}
Bourdieu, P.; and Wacquant, L.~J. 1992.
\newblock \emph{An invitation to reflexive sociology}.
\newblock University of Chicago press.

\bibitem[{Boyon(2022)}]{boyon2022opinions}
Boyon, N. 2022.
\newblock Opinions about AI vary depending on countries’ level of economic development.

\bibitem[{Boyon(2023)}]{boyon2023nervous}
Boyon, N. 2023.
\newblock AI is making the world more nervous.

\bibitem[{Brown(2023)}]{Brown2023}
Brown, M. 2023.
\newblock How to Run Surveys at Every Stage of the Design Cycle.

\bibitem[{Byun, Vasicek, and Seppi(2023)}]{byun2023dispensing}
Byun, C.; Vasicek, P.; and Seppi, K. 2023.
\newblock Dispensing with Humans in Human-Computer Interaction Research.
\newblock In \emph{Extended Abstracts of the 2023 CHI Conference on Human Factors in Computing Systems}. ACM.

\bibitem[{Cannell et~al.(1977)Cannell, Marquis, Laurent et~al.}]{cannell1977summary}
Cannell, C.~F.; Marquis, K.~H.; Laurent, A.; et~al. 1977.
\newblock \emph{A summary of studies of interviewing methodology}.
\newblock US Government Printing Office, Washington, DC 20402.

\bibitem[{Cave, Coughlan, and Dihal(2019)}]{cave2019scary}
Cave, S.; Coughlan, K.; and Dihal, K. 2019.
\newblock ``Scary Robots'': Examining Public Responses to AI.
\newblock In \emph{Proceedings of the 2019 AAAI/ACM Conference on AI, Ethics, and Society}. ACM.

\bibitem[{Center(2024)}]{PewSurveys}
Center, P.~R. 2024.
\newblock Writing Survey Questions.

\bibitem[{Chasalow and Levy(2021)}]{chasalow2021representativeness}
Chasalow, K.; and Levy, K. 2021.
\newblock Representativeness in Statistics, Politics, and Machine Learning.
\newblock In \emph{Proceedings of the 2021 {ACM} {Conference} on {Fairness}, {Accountability}, and {Transparency}}. ACM.

\bibitem[{Chilisa(2019)}]{chilisa2019indigenous}
Chilisa, B. 2019.
\newblock \emph{Indigenous research methodologies}.
\newblock Sage publications.

\bibitem[{Chmielinski et~al.(2022)Chmielinski, Newman, Taylor, Joseph, Thomas, Yurkofsky, and Qiu}]{chmielinski2022dataset}
Chmielinski, K.~S.; Newman, S.; Taylor, M.; Joseph, J.; Thomas, K.; Yurkofsky, J.; and Qiu, Y.~C. 2022.
\newblock The Dataset Nutrition Label (2nd Gen): Leveraging Context to Mitigate Harms in Artificial Intelligence.

\bibitem[{Choi and Pak(2005)}]{choi2005catalog}
Choi, B. C.~K.; and Pak, A. W.~P. 2005.
\newblock A catalog of biases in questionnaires.
\newblock \emph{Prev. Chronic Dis.}

\bibitem[{Colton and Covert(2007)}]{colton2007designing}
Colton, D.; and Covert, R.~W. 2007.
\newblock \emph{Designing and constructing instruments for social research and evaluation}.
\newblock John Wiley \& Sons.

\bibitem[{Comber et~al.(2020)Comber, Bardzell, Bardzell, Hazas, and Muller}]{comber2020announcing}
Comber, R.; Bardzell, S.; Bardzell, J.; Hazas, M.; and Muller, M. 2020.
\newblock Announcing a new CHI subcommittee: critical and sustainable computing.
\newblock \emph{Interactions}.

\bibitem[{Constantinides et~al.(2024{\natexlab{a}})Constantinides, Bogucka, Quercia, Kallio, and Tahaei}]{constantinides2024method}
Constantinides, M.; Bogucka, E.; Quercia, D.; Kallio, S.; and Tahaei, M. 2024{\natexlab{a}}.
\newblock A Method for Generating Dynamic Responsible AI Guidelines for Collaborative Action.
\newblock \emph{Conference on Computer-Supported Cooperative Work and Social Computing (CSCW)}.

\bibitem[{Constantinides et~al.(2024{\natexlab{b}})Constantinides, Tahaei, Quercia, Stumpf, Madaio, Kennedy, Wilcox, Vitak, Cramer, Bogucka, Baeza-Yates, Luger, Holbrook, Muller, Blumenfeld, and Pistilli}]{constantinides2024role}
Constantinides, M.; Tahaei, M.; Quercia, D.; Stumpf, S.; Madaio, M.; Kennedy, S.; Wilcox, L.; Vitak, J.; Cramer, H.; Bogucka, E.~P.; Baeza-Yates, R.; Luger, E.; Holbrook, J.; Muller, M.; Blumenfeld, I.~G.; and Pistilli, G. 2024{\natexlab{b}}.
\newblock Implications of Regulations on the Use of AI and Generative AI for Human-Centered Responsible Artificial Intelligence.
\newblock In \emph{Extended Abstracts of the 2024 CHI Conference on Human Factors in Computing Systems}. ACM.

\bibitem[{Converse and Presser(1986)}]{converse1986survey}
Converse, J.~M.; and Presser, S. 1986.
\newblock \emph{Survey questions: Handcrafting the standardized questionnaire}.
\newblock Sage.

\bibitem[{Cooper et~al.(2022)Cooper, Horne, Hayes, Heldreth, Lahav, Holbrook, and Wilcox}]{cooper2022systematic}
Cooper, N.; Horne, T.; Hayes, G.~R.; Heldreth, C.; Lahav, M.; Holbrook, J.; and Wilcox, L. 2022.
\newblock A Systematic Review and Thematic Analysis of Community-Collaborative Approaches to Computing Research.
\newblock In \emph{Proceedings of the 2022 CHI Conference on Human Factors in Computing Systems}. ACM.

\bibitem[{Crabtree et~al.(2009)Crabtree, Rodden, Tolmie, and Button}]{crabtree2009ethnography}
Crabtree, A.; Rodden, T.; Tolmie, P.; and Button, G. 2009.
\newblock Ethnography Considered Harmful.
\newblock In \emph{Proceedings of the SIGCHI Conference on Human Factors in Computing Systems}. ACM.

\bibitem[{Crawford and Schultz(2014)}]{crawford2014big}
Crawford, K.; and Schultz, J. 2014.
\newblock Big data and due process: Toward a framework to redress predictive privacy harms.
\newblock \emph{BCL Rev.}

\bibitem[{Crisan et~al.(2022)Crisan, Drouhard, Vig, and Rajani}]{crisan2022interactive}
Crisan, A.; Drouhard, M.; Vig, J.; and Rajani, N. 2022.
\newblock Interactive Model Cards: A Human-Centered Approach to Model Documentation.
\newblock In \emph{2022 {ACM} {Conference} on {Fairness}, {Accountability}, and {Transparency}}. ACM.

\bibitem[{Cunliffe(2004)}]{cunliffe2004on}
Cunliffe, A.~L. 2004.
\newblock On Becoming a Critically Reflexive Practitioner.
\newblock \emph{Journal of Management Education}.

\bibitem[{Davani et~al.(2024)Davani, D{\'\i}az, Baker, and Prabhakaran}]{davani2024disentangling}
Davani, A.; D{\'\i}az, M.; Baker, D.; and Prabhakaran, V. 2024.
\newblock Disentangling Perceptions of Offensiveness: Cultural and Moral Correlates.
\newblock In \emph{The 2024 ACM Conference on Fairness, Accountability, and Transparency}.

\bibitem[{Delgado et~al.(2023)Delgado, Yang, Madaio, and Yang}]{delgado2023participatory}
Delgado, F.; Yang, S.; Madaio, M.; and Yang, Q. 2023.
\newblock The Participatory Turn in AI Design: Theoretical Foundations and the Current State of Practice.
\newblock In \emph{Proceedings of the 3rd ACM Conference on Equity and Access in Algorithms, Mechanisms, and Optimization}. ACM.

\bibitem[{Dennler et~al.(2023)Dennler, Ovalle, Singh, Soldaini, Subramonian, Tu, Agnew, Ghosh, Yee, Peradejordi, Talat, Russo, and Pinhal}]{dennler2023bound}
Dennler, N.; Ovalle, A.; Singh, A.; Soldaini, L.; Subramonian, A.; Tu, H.; Agnew, W.; Ghosh, A.; Yee, K.; Peradejordi, I.~F.; Talat, Z.; Russo, M.; and Pinhal, J. D. J. D.~P. 2023.
\newblock Bound by the Bounty: Collaboratively Shaping Evaluation Processes for Queer AI Harms.
\newblock In \emph{Proceedings of the 2023 AAAI/ACM Conference on AI, Ethics, and Society}. ACM.

\bibitem[{Denzin, Lincoln, and Smith(2008)}]{denzin2008handbook}
Denzin, N.~K.; Lincoln, Y.~S.; and Smith, L.~T. 2008.
\newblock \emph{Handbook of critical and indigenous methodologies}.
\newblock Sage.

\bibitem[{D\'{\i}az et~al.(2022)D\'{\i}az, Kivlichan, Rosen, Baker, Amironesei, Prabhakaran, and Denton}]{CrowdSheets}
D\'{\i}az, M.; Kivlichan, I.; Rosen, R.; Baker, D.; Amironesei, R.; Prabhakaran, V.; and Denton, E. 2022.
\newblock CrowdWorkSheets: Accounting for Individual and Collective Identities Underlying Crowdsourced Dataset Annotation.
\newblock In \emph{2022 ACM Conference on Fairness, Accountability, and Transparency}. Association for Computing Machinery.

\bibitem[{D'Ignazio and Klein(2020)}]{d_ignazio2020data}
D'Ignazio, C.; and Klein, L.~F. 2020.
\newblock \emph{Data feminism}.
\newblock MIT press.

\bibitem[{Dijkstra(1968)}]{dijkstra1968harmful}
Dijkstra, E.~W. 1968.
\newblock Letters to the Editor: Go to Statement Considered Harmful.
\newblock \emph{Commun. ACM}.

\bibitem[{Dillman, Tortora, and Bowker(1998)}]{dillman1998principles}
Dillman, D.~A.; Tortora, R.~D.; and Bowker, D. 1998.
\newblock Principles for constructing web surveys.
\newblock In \emph{Joint Meetings of the American Statistical Association}.

\bibitem[{Dillman et~al.(1978)}]{dillman1978mail}
Dillman, D.~A.; et~al. 1978.
\newblock \emph{Mail and telephone surveys: The total design method}.
\newblock Wiley New York.

\bibitem[{Do et~al.(2024)Do, De~Los~Santos, Muller, and Savage}]{do2024gigsousveillance}
Do, K.; De~Los~Santos, M.; Muller, M.; and Savage, S. 2024.
\newblock GigSousveillance: Designing Gig Worker Centric Sousveillance Tools.
\newblock In \emph{ACM CHI Conference on Human Factors in Computing Systems}.

\bibitem[{Doherty and Doherty(2018)}]{doherty2018engagement}
Doherty, K.; and Doherty, G. 2018.
\newblock Engagement in HCI: Conception, Theory and Measurement.
\newblock \emph{ACM Comput. Surv.}

\bibitem[{Douglas, Ewell, and Brauer(2023)}]{douglas2023data}
Douglas, B.~D.; Ewell, P.~J.; and Brauer, M. 2023.
\newblock Data quality in online human-subjects research: Comparisons between MTurk, Prolific, CloudResearch, Qualtrics, and SONA.
\newblock \emph{Plos one}.

\bibitem[{Douglas and Wildavsky(1983)}]{douglas1983risk}
Douglas, M.; and Wildavsky, A. 1983.
\newblock \emph{Risk and culture: An essay on the selection of technological and environmental dangers}.
\newblock Univ of California Press.

\bibitem[{Dourish et~al.(2020)Dourish, Lawrence, Leong, and Wadley}]{dourish2020on}
Dourish, P.; Lawrence, C.; Leong, T.~W.; and Wadley, G. 2020.
\newblock On Being Iterated: The Affective Demands of Design Participation.
\newblock In \emph{Proceedings of the 2020 CHI Conference on Human Factors in Computing Systems}. ACM.

\bibitem[{Dreber and Johannesson(2019)}]{dreber2019statistical}
Dreber, A.; and Johannesson, M. 2019.
\newblock Statistical significance and the replication crisis in the social sciences.
\newblock In \emph{Oxford research encyclopedia of economics and finance}.

\bibitem[{Duarte(2021)}]{duarte2021native}
Duarte, M.~E. 2021.
\newblock Native and indigenous women’s cyber-defense of lands and peoples.
\newblock \emph{Networked Feminisms: Activist Assemblies and Digital Practices}.

\bibitem[{Dugan et~al.(2021)Dugan, Namazi, Cavallari, Rinker, Preston, Steele, and Cherniack}]{dugan2021participatory}
Dugan, A.~G.; Namazi, S.; Cavallari, J.~M.; Rinker, R.~D.; Preston, J.~C.; Steele, V.~L.; and Cherniack, M.~G. 2021.
\newblock Participatory survey design of a workforce health needs assessment for correctional supervisors.
\newblock \emph{American journal of industrial medicine}.

\bibitem[{Echtler and H\"{a}u\ss{}ler(2018)}]{echtler2018open}
Echtler, F.; and H\"{a}u\ss{}ler, M. 2018.
\newblock Open Source, Open Science, and the Replication Crisis in HCI.
\newblock In \emph{Extended Abstracts of the 2018 CHI Conference on Human Factors in Computing Systems}. ACM.

\bibitem[{Epstein et~al.(2023)Epstein, Bordyug, Chen, Chen, Ginther, Kirkish, and Stead}]{epstein2023toward}
Epstein, R.; Bordyug, M.; Chen, Y.-H.; Chen, Y.; Ginther, A.; Kirkish, G.; and Stead, H. 2023.
\newblock Toward the search for the perfect blade runner: a large-scale, international assessment of a test that screens for ``humanness sensitivity''.
\newblock \emph{AI \& SOCIETY}.

\bibitem[{Epstein et~al.(2013)Epstein, Yanovich, Moran, and Heled}]{epstein2013physiological}
Epstein, Y.; Yanovich, R.; Moran, D.~S.; and Heled, Y. 2013.
\newblock Physiological employment standards IV: integration of women in combat units physiological and medical considerations.
\newblock \emph{European journal of applied physiology}.

\bibitem[{Etienne and Cova(2024)}]{etienne2024autonomous}
Etienne, H.; and Cova, F. 2024.
\newblock The more they think, the less they want: studying people's attitudes about autonomous vehicles could also contribute to shaping them.
\newblock \emph{AI and Ethics}.

\bibitem[{Fast and Horvitz(2017)}]{fast2017long}
Fast, E.; and Horvitz, E. 2017.
\newblock Long-Term Trends in the Public Perception of Artificial Intelligence.
\newblock In \emph{Proceedings of the Thirty-First AAAI Conference on Artificial Intelligence}. AAAI Press.

\bibitem[{Feffer et~al.(2023)Feffer, Skirpan, Lipton, and Heidari}]{feffer2023from}
Feffer, M.; Skirpan, M.; Lipton, Z.; and Heidari, H. 2023.
\newblock From Preference Elicitation to Participatory ML: A Critical Survey \& Guidelines for Future Research.
\newblock In \emph{Proceedings of the 2023 AAAI/ACM Conference on AI, Ethics, and Society}. ACM.

\bibitem[{Feinberg(2017)}]{feinberg2017design}
Feinberg, M. 2017.
\newblock A Design Perspective on Data.
\newblock In \emph{Proceedings of the 2017 CHI Conference on Human Factors in Computing Systems}. ACM.

\bibitem[{Feinberg(2022)}]{feinberg2022everyday}
Feinberg, M. 2022.
\newblock \emph{Everyday Adventures with Unruly Data}.
\newblock MIT Press.

\bibitem[{{Finance Center for South-South Cooperation}(2024)}]{uni2024south}
{Finance Center for South-South Cooperation}. 2024.
\newblock Global South Countries.

\bibitem[{Fink(2003)}]{fink2003design}
Fink, A. 2003.
\newblock \emph{How to design survey studies}.
\newblock Sage.

\bibitem[{Flicker et~al.(2010)Flicker, Guta, Larkin, Flynn, Fridkin, Travers, Pole, and Layne}]{flicker2010survey}
Flicker, S.; Guta, A.; Larkin, J.; Flynn, S.; Fridkin, A.; Travers, R.; Pole, J.~D.; and Layne, C. 2010.
\newblock Survey design from the ground up: Collaboratively creating the Toronto Teen Survey.
\newblock \emph{Health Promotion Practice}.

\bibitem[{Fowler(2013)}]{fowler2013survey}
Fowler, F.~J. 2013.
\newblock \emph{Survey Research Methods}.
\newblock SAGE Publications.

\bibitem[{Fowler~Jr and Mangione(1990)}]{fowler1990standardized}
Fowler~Jr, F.~J.; and Mangione, T.~W. 1990.
\newblock \emph{Standardized survey interviewing: Minimizing interviewer-related error}.
\newblock Sage.

\bibitem[{Francis et~al.(2010)Francis, Johnston, Robertson, Glidewell, Entwistle, Eccles, and Grimshaw}]{francis2010what}
Francis, J.~J.; Johnston, M.; Robertson, C.; Glidewell, L.; Entwistle, V.; Eccles, M.~P.; and Grimshaw, J.~M. 2010.
\newblock What is an adequate sample size? Operationalising data saturation for theory-based interview studies.
\newblock \emph{Psychology \& Health}.

\bibitem[{Franken and Wattenberg(2019)}]{franken2019impact}
Franken, S.; and Wattenberg, M. 2019.
\newblock The impact of AI on employment and organisation in the industrial working environment of the future.
\newblock In \emph{ECIAIR 2019 European Conference on the Impact of Artificial Intelligence and Robotics}. Academic Conferences and publishing limited.

\bibitem[{Freese and Peterson(2017)}]{freese2017replication}
Freese, J.; and Peterson, D. 2017.
\newblock Replication in social science.
\newblock \emph{Annual Review of Sociology}.

\bibitem[{Gilovich, Keltner, and Nisbett(2006)}]{gilovich2006being}
Gilovich, T.; Keltner, D.; and Nisbett, R.~E. 2006.
\newblock Being a member of a stigmatized group: stereotype threat.
\newblock \emph{Gilovich, Thomas; Keltner, Dacher; Nisbett, Richard E., Social psychology, New York: WW Norton}.

\bibitem[{Gomez~Ortega et~al.(2023)Gomez~Ortega, Bourgeois, Hutiri, and Kortuem}]{gomez2023beyond}
Gomez~Ortega, A.; Bourgeois, J.; Hutiri, W.~T.; and Kortuem, G. 2023.
\newblock Beyond data transactions: a framework for meaningfully informed data donation.
\newblock \emph{AI \& SOCIETY}.

\bibitem[{Google(2024)}]{google2024global}
Google. 2024.
\newblock Global Study Shows Optimism About AI's Potential.

\bibitem[{Greenberg and Buxton(2008)}]{greenberg2008usability}
Greenberg, S.; and Buxton, B. 2008.
\newblock Usability Evaluation Considered Harmful (Some of the Time).
\newblock In \emph{Proceedings of the SIGCHI Conference on Human Factors in Computing Systems}. ACM.

\bibitem[{Groves et~al.(2023)Groves, Peppin, Strait, and Brennan}]{groves2023goingpublicpaper}
Groves, L.; Peppin, A.; Strait, A.; and Brennan, J. 2023.
\newblock Going Public: The Role of Public Participation Approaches in Commercial AI Labs.
\newblock In \emph{Proceedings of the 2023 ACM Conference on Fairness, Accountability, and Transparency}. ACM.

\bibitem[{Groves et~al.(2009)Groves, Fowler~Jr, Couper, Lepkowski, Singer, and Tourangeau}]{groves2009survey}
Groves, R.~M.; Fowler~Jr, F.~J.; Couper, M.~P.; Lepkowski, J.~M.; Singer, E.; and Tourangeau, R. 2009.
\newblock \emph{Survey methodology}.
\newblock John Wiley \& Sons.

\bibitem[{Groves et~al.(2011)Groves, Fowler~Jr, Couper, Lepkowski, Singer, and Tourangeau}]{groves2011survey}
Groves, R.~M.; Fowler~Jr, F.~J.; Couper, M.~P.; Lepkowski, J.~M.; Singer, E.; and Tourangeau, R. 2011.
\newblock \emph{Survey methodology}.
\newblock John Wiley \& Sons.

\bibitem[{Guest, Bunce, and Johnson(2006)}]{guest2006how}
Guest, G.; Bunce, A.; and Johnson, L. 2006.
\newblock How Many Interviews Are Enough?: An Experiment with Data Saturation and Variability.
\newblock \emph{Field Methods}.

\bibitem[{Hansen, Fourie, and Meyer(2021)}]{hansen2021third}
Hansen, P.; Fourie, I.; and Meyer, A. 2021.
\newblock \emph{Third space, information sharing, and participatory design}.
\newblock Springer.

\bibitem[{Hara et~al.(2018)Hara, Adams, Milland, Savage, Callison-Burch, and Bigham}]{hara2018data}
Hara, K.; Adams, A.; Milland, K.; Savage, S.; Callison-Burch, C.; and Bigham, J.~P. 2018.
\newblock A Data-Driven Analysis of Workers' Earnings on Amazon Mechanical Turk.
\newblock In \emph{Proceedings of the 2018 CHI Conference on Human Factors in Computing Systems}. ACM.

\bibitem[{Havens et~al.(2020)Havens, Terras, Bach, and Alex}]{havens2020situated}
Havens, L.; Terras, M.; Bach, B.; and Alex, B. 2020.
\newblock Situated Data, Situated Systems: A Methodology to Engage with Power Relations in Natural Language Processing Research.
\newblock In \emph{Proceedings of the Second Workshop on Gender Bias in Natural Language Processing}. Association for Computational Linguistics.

\bibitem[{Hayes(2014)}]{hayes2014knowing}
Hayes, G.~R. 2014.
\newblock Knowing by doing: action research as an approach to HCI.
\newblock In \emph{Ways of Knowing in HCI}. Springer.

\bibitem[{Hertzberg, Liberti, and Paravisini(2010)}]{hertzberg2010information}
Hertzberg, A.; Liberti, J.~M.; and Paravisini, D. 2010.
\newblock Information and incentives inside the firm: Evidence from loan officer rotation.
\newblock \emph{The Journal of Finance}.

\bibitem[{Himmelreich(2023)}]{himmelreich2023against}
Himmelreich, J. 2023.
\newblock Against ``Democratizing {AI}''.
\newblock \emph{AI \& SOCIETY}.

\bibitem[{Holdren, Sunstein, and Siddiqui(2011)}]{holdren2011principles}
Holdren, J.~P.; Sunstein, C.~R.; and Siddiqui, I.~A. 2011.
\newblock \emph{Principles for regulation and oversight of emerging technologies}.
\newblock Office of Science and Technology Policy.

\bibitem[{Hosseini, Resnik, and Holmes(2023)}]{hosseini2023ethics}
Hosseini, M.; Resnik, D.~B.; and Holmes, K. 2023.
\newblock The ethics of disclosing the use of artificial intelligence tools in writing scholarly manuscripts.
\newblock \emph{Research Ethics}.

\bibitem[{Houser(2019)}]{houser2019can}
Houser, K.~A. 2019.
\newblock Can AI solve the diversity problem in the tech industry: Mitigating noise and bias in employment decision-making.
\newblock \emph{Stan. Tech. L. Rev.}

\bibitem[{Huang et~al.(2024)Huang, Siddarth, Lovitt, Liao, Durmus, Tamkin, and Ganguli}]{huang2024collective}
Huang, S.; Siddarth, D.; Lovitt, L.; Liao, T.~I.; Durmus, E.; Tamkin, A.; and Ganguli, D. 2024.
\newblock Collective Constitutional AI: Aligning a Language Model with Public Input.
\newblock In \emph{The 2024 ACM Conference on Fairness, Accountability, and Transparency}.

\bibitem[{Hursthouse and Pettigrove(2023)}]{sep-ethics-virtue}
Hursthouse, R.; and Pettigrove, G. 2023.
\newblock Virtue Ethics.
\newblock In Zalta, E.~N.; and Nodelman, U., eds., \emph{The {Stanford} Encyclopedia of Philosophy}. Metaphysics Research Lab, Stanford University, {F}all 2023 edition.

\bibitem[{Ikkatai et~al.(2023)Ikkatai, Hartwig, Takanashi, and Yokoyama}]{ikkatai2023segmentation}
Ikkatai, Y.; Hartwig, T.; Takanashi, N.; and Yokoyama, H.~M. 2023.
\newblock Segmentation of ethics, legal, and social issues (ELSI) related to AI in Japan, the United States, and Germany.
\newblock \emph{AI and Ethics}.

\bibitem[{Iliadis and Russo(2016)}]{iliadis2016critical}
Iliadis, A.; and Russo, F. 2016.
\newblock Critical data studies: An introduction.
\newblock \emph{Big Data \& Society}.

\bibitem[{Irani et~al.(2010)Irani, Vertesi, Dourish, Philip, and Grinter}]{irani2010postcolonial}
Irani, L.; Vertesi, J.; Dourish, P.; Philip, K.; and Grinter, R.~E. 2010.
\newblock Postcolonial Computing: A Lens on Design and Development.
\newblock In \emph{Proceedings of the SIGCHI Conference on Human Factors in Computing Systems}. ACM.

\bibitem[{Jakesch et~al.(2022)Jakesch, Buçinca, Amershi, and Olteanu}]{jakesch2022how}
Jakesch, M.; Buçinca, Z.; Amershi, S.; and Olteanu, A. 2022.
\newblock How Different Groups Prioritize Ethical Values for Responsible AI.
\newblock In \emph{2022 {ACM} {Conference} on {Fairness}, {Accountability}, and {Transparency}}. ACM.

\bibitem[{Jamieson, Govaart, and Pownall(2023)}]{jamieson2023reflexivity}
Jamieson, M.~K.; Govaart, G.~H.; and Pownall, M. 2023.
\newblock Reflexivity in quantitative research: A rationale and beginner's guide.
\newblock \emph{Social and Personality Psychology Compass}.

\bibitem[{Kapania et~al.(2022)Kapania, Siy, Clapper, SP, and Sambasivan}]{kapania2022because}
Kapania, S.; Siy, O.; Clapper, G.; SP, A.~M.; and Sambasivan, N. 2022.
\newblock ''Because AI is 100\% Right and Safe'': User Attitudes and Sources of AI Authority in India.
\newblock In \emph{Proceedings of the 2022 {CHI} {Conference} on {Human} {Factors} in {Computing} {Systems}}. ACM.

\bibitem[{Karsgaard(2023)}]{karsgaard2023new}
Karsgaard, C. 2023.
\newblock New Visions for Anti-colonial Digital Methods.
\newblock In \emph{Instagram as Public Pedagogy: Online Activism and the Trans Mountain Pipeline}. Springer.

\bibitem[{Kaufmann, Schulze, and Veit(2011)}]{kaufmann2011more}
Kaufmann, N.; Schulze, T.; and Veit, D. 2011.
\newblock More than fun and money. worker motivation in crowdsourcing--a study on mechanical turk.

\bibitem[{Kelley et~al.(2003)Kelley, Clark, Brown, and Sitzia}]{kelley2003good}
Kelley, K.; Clark, B.; Brown, V.; and Sitzia, J. 2003.
\newblock Good practice in the conduct and reporting of survey research.
\newblock \emph{International Journal for Quality in health care}.

\bibitem[{Kelley et~al.(2021)Kelley, Yang, Heldreth, Moessner, Sedley, Kramm, Newman, and Woodruff}]{kelley2021exciting}
Kelley, P.~G.; Yang, Y.; Heldreth, C.; Moessner, C.; Sedley, A.; Kramm, A.; Newman, D.~T.; and Woodruff, A. 2021.
\newblock Exciting, Useful, Worrying, Futuristic: Public Perception of Artificial Intelligence in 8 Countries.
\newblock In \emph{Proceedings of the 2021 AAAI/ACM Conference on AI, Ethics, and Society}. ACM.

\bibitem[{Kieslich and L{\"u}nich(2024)}]{kieslich2024regulating}
Kieslich, K.; and L{\"u}nich, M. 2024.
\newblock Regulating AI-Based Remote Biometric Identification. Investigating the Public Demand for Bans, Audits, and Public Database Registrations.
\newblock In \emph{The 2024 ACM Conference on Fairness, Accountability, and Transparency}.

\bibitem[{Kitchin and Lauriault(2014)}]{kitchin2014towards}
Kitchin, R.; and Lauriault, T. 2014.
\newblock Towards critical data studies: Charting and unpacking data assemblages and their work.

\bibitem[{Ko et~al.(2023)Ko, Beitlers, Wortzman, Davidson, Oleson, Kirdani-Ryan, Druga, and Everson}]{ko2023critically}
Ko, A.~J.; Beitlers, A.; Wortzman, B.; Davidson, M.; Oleson, A.; Kirdani-Ryan, M.; Druga, S.; and Everson, J. 2023.
\newblock \emph{Critically Conscious Computing: Methods for Secondary Education}.

\bibitem[{Kovach(2021)}]{kovach2021indigenous}
Kovach, M. 2021.
\newblock \emph{Indigenous methodologies: Characteristics, conversations, and contexts}.
\newblock University of Toronto press.

\bibitem[{Kramer et~al.(2018)Kramer, Schaich~Borg, Conitzer, and Sinnott-Armstrong}]{kramer2018when}
Kramer, M.~F.; Schaich~Borg, J.; Conitzer, V.; and Sinnott-Armstrong, W. 2018.
\newblock When Do People Want AI to Make Decisions?
\newblock In \emph{Proceedings of the 2018 AAAI/ACM Conference on AI, Ethics, and Society}. Association for Computing Machinery.

\bibitem[{Krosnick(1999)}]{krosnick1999survey}
Krosnick, J.~A. 1999.
\newblock Survey Research.
\newblock \emph{Annual Review of Psychology}.

\bibitem[{Kuhn, Parker, and Lefthand-Begay(2020)}]{kuhn2020indigenous}
Kuhn, N.~S.; Parker, M.; and Lefthand-Begay, C. 2020.
\newblock Indigenous research ethics requirements: an examination of six tribal institutional review board applications and processes in the United States.
\newblock \emph{Journal of empirical research on human research ethics}.

\bibitem[{Kwet(2019)}]{kwet2019digial}
Kwet, M. 2019.
\newblock Digital colonialism: US empire and the new imperialism in the Global South.
\newblock \emph{Race \& Class}.

\bibitem[{Laufer et~al.(2022)Laufer, Jain, Cooper, Kleinberg, and Heidari}]{laufer2022four}
Laufer, B.; Jain, S.; Cooper, A.~F.; Kleinberg, J.; and Heidari, H. 2022.
\newblock Four Years of FAccT: A Reflexive, Mixed-Methods Analysis of Research Contributions, Shortcomings, and Future Prospects.
\newblock In \emph{2022 {ACM} {Conference} on {Fairness}, {Accountability}, and {Transparency}}. ACM.

\bibitem[{Lee et~al.(2002)Lee, Jones, Mineyama, and Zhang}]{lee2002cultural}
Lee, J.~W.; Jones, P.~S.; Mineyama, Y.; and Zhang, X.~E. 2002.
\newblock Cultural differences in responses to a Likert scale.
\newblock \emph{Research in nursing \& health}.

\bibitem[{Linxen et~al.(2021)Linxen, Sturm, Br\"{u}hlmann, Cassau, Opwis, and Reinecke}]{linxen2021how}
Linxen, S.; Sturm, C.; Br\"{u}hlmann, F.; Cassau, V.; Opwis, K.; and Reinecke, K. 2021.
\newblock How WEIRD is CHI?
\newblock In \emph{Proceedings of the 2021 CHI Conference on Human Factors in Computing Systems}. ACM.

\bibitem[{Loefflad and Grossklags(2024)}]{loefflad2024types}
Loefflad, C.; and Grossklags, J. 2024.
\newblock How the Types of Consequences in Social Scoring Systems Shape People's Perceptions and Behavioral Reactions.
\newblock In \emph{The 2024 ACM Conference on Fairness, Accountability, and Transparency}.

\bibitem[{Lovett et~al.(2018)Lovett, Bajaba, Lovett, and Simmering}]{lovett2018data}
Lovett, M.; Bajaba, S.; Lovett, M.; and Simmering, M.~J. 2018.
\newblock Data quality from crowdsourced surveys: A mixed method inquiry into perceptions of Amazon's Mechanical Turk Masters.
\newblock \emph{Applied Psychology}.

\bibitem[{Lu et~al.(2024)Lu, Moy, Ackerman, Morenoff, and Dillahunt}]{lu2024perceptions}
Lu, A.~J.; Moy, C.; Ackerman, M.~S.; Morenoff, J.; and Dillahunt, T.~R. 2024.
\newblock Perceptions of Policing Surveillance Technologies in Detroit: Moving Beyond "Better than Nothing".
\newblock In \emph{Proceedings of the 2024 ACM Conference on Fairness, Accountability, and Transparency}. Association for Computing Machinery.

\bibitem[{L{\"u}nich and Keller(2024)}]{lunich2024explainable}
L{\"u}nich, M.; and Keller, B. 2024.
\newblock Explainable Artificial Intelligence for Academic Performance Prediction. An Experimental Study on the Impact of Accuracy and Simplicity of Decision Trees on Causability and Fairness Perceptions.
\newblock In \emph{The 2024 ACM Conference on Fairness, Accountability, and Transparency}.

\bibitem[{Mann and Daly(2019)}]{mann2019big}
Mann, M.; and Daly, A. 2019.
\newblock (Big) Data and the North-in-South: Australia’s Informational Imperialism and Digital Colonialism.
\newblock \emph{Television \& New Media}.

\bibitem[{Mao et~al.(2019)Mao, Wang, Muller, Varshney, Baldini, Dugan, and Mojsilovi{\'c}}]{mao2019data}
Mao, Y.; Wang, D.; Muller, M.; Varshney, K.~R.; Baldini, I.; Dugan, C.; and Mojsilovi{\'c}, A. 2019.
\newblock How data scientistswork together with domain experts in scientific collaborations: To find the right answer or to ask the right question?
\newblock \emph{Proceedings of the ACM on Human-Computer Interaction}.

\bibitem[{Marenko(2018)}]{marenko2018futurecrafting}
Marenko, B. 2018.
\newblock Futurecrafting. A speculative method for an imaginative AI.
\newblock \emph{AAAI Spring Symposium Series}.

\bibitem[{McKee(2023)}]{mckee2023human}
McKee, K.~R. 2023.
\newblock Human participants in AI research: Ethics and transparency in practice.

\bibitem[{Miceli et~al.(2021)Miceli, Yang, Naudts, Schuessler, Serbanescu, and Hanna}]{miceli2021documenting}
Miceli, M.; Yang, T.; Naudts, L.; Schuessler, M.; Serbanescu, D.; and Hanna, A. 2021.
\newblock Documenting Computer Vision Datasets: An Invitation to Reflexive Data Practices.
\newblock In \emph{Proceedings of the 2021 {ACM} {Conference} on {Fairness}, {Accountability}, and {Transparency}}. ACM.

\bibitem[{Midena and Yeo(2022)}]{midena2022towards}
Midena, D.; and Yeo, R. 2022.
\newblock Towards a history of the questionnaire.

\bibitem[{Mills(2023)}]{mills2023sociological}
Mills, C.~W. 2023.
\newblock The sociological imagination.
\newblock In \emph{Social Work}. Routledge.

\bibitem[{Mir et~al.(2012)Mir, Salway, Kai, Karlsen, Bhopal, Ellison, and Sheikh}]{mir2012principles}
Mir, G.; Salway, S.; Kai, J.; Karlsen, S.; Bhopal, R.; Ellison, G.~T.; and Sheikh, A. 2012.
\newblock {Principles for research on ethnicity and health: the Leeds Consensus Statement}.
\newblock \emph{European Journal of Public Health}.

\bibitem[{Mitchell et~al.(2019)Mitchell, Wu, Zaldivar, Barnes, Vasserman, Hutchinson, Spitzer, Raji, and Gebru}]{mitchell2019model}
Mitchell, M.; Wu, S.; Zaldivar, A.; Barnes, P.; Vasserman, L.; Hutchinson, B.; Spitzer, E.; Raji, I.~D.; and Gebru, T. 2019.
\newblock Model Cards for Model Reporting.
\newblock In \emph{Proceedings of the {Conference} on {Fairness}, {Accountability}, and {Transparency}}. ACM.

\bibitem[{Moloi and Marwala(2021)}]{moloi2021artificial}
Moloi, T.; and Marwala, T. 2021.
\newblock \emph{Artificial Intelligence and the Changing Nature of Corporations: How Technologies Shape Strategy and Operations}.
\newblock Springer Nature.

\bibitem[{Morley et~al.(2020)Morley, Machado, Burr, Cowls, Joshi, Taddeo, and Floridi}]{morley2020ethics}
Morley, J.; Machado, C.~C.; Burr, C.; Cowls, J.; Joshi, I.; Taddeo, M.; and Floridi, L. 2020.
\newblock The ethics of AI in health care: a mapping review.
\newblock \emph{Social Science \& Medicine}.

\bibitem[{Muller et~al.(2016)Muller, Guha, Baumer, Mimno, and Shami}]{muller2016machine}
Muller, M.; Guha, S.; Baumer, E.~P.; Mimno, D.; and Shami, N.~S. 2016.
\newblock Machine learning and grounded theory method: convergence, divergence, and combination.
\newblock In \emph{Proceedings of the 2016 ACM International Conference on Supporting Group Work}.

\bibitem[{Muller et~al.(2019)Muller, Lange, Wang, Piorkowski, Tsay, Liao, Dugan, and Erickson}]{muller2019data}
Muller, M.; Lange, I.; Wang, D.; Piorkowski, D.; Tsay, J.; Liao, Q.~V.; Dugan, C.; and Erickson, T. 2019.
\newblock How Data Science Workers Work with Data: Discovery, Capture, Curation, Design, Creation.
\newblock In \emph{Proceedings of the 2019 CHI Conference on Human Factors in Computing Systems}. ACM.

\bibitem[{Muller and Strohmayer(2022)}]{muller2022forgetting}
Muller, M.; and Strohmayer, A. 2022.
\newblock Forgetting Practices in the Data Sciences.
\newblock In \emph{Proceedings of the 2022 {CHI} {Conference} on {Human} {Factors} in {Computing} {Systems}}. ACM.

\bibitem[{Muller et~al.(2021)Muller, Wolf, Andres, Desmond, Joshi, Ashktorab, Sharma, Brimijoin, Pan, Duesterwald, and Dugan}]{muller2021designing}
Muller, M.; Wolf, C.~T.; Andres, J.; Desmond, M.; Joshi, N.~N.; Ashktorab, Z.; Sharma, A.; Brimijoin, K.; Pan, Q.; Duesterwald, E.; and Dugan, C. 2021.
\newblock Designing Ground Truth and the Social Life of Labels.
\newblock In \emph{Proceedings of the 2021 CHI Conference on Human Factors in Computing Systems}. ACM.

\bibitem[{Muller(1997)}]{muller1997ethnocritical}
Muller, M.~J. 1997.
\newblock Ethnocritical heuristics for reflecting on work with users and other interested parties.
\newblock In \emph{Computers and design in context}.

\bibitem[{Muller and Kuhn(1993)}]{muller1993participatory}
Muller, M.~J.; and Kuhn, S. 1993.
\newblock Participatory design.
\newblock \emph{Communications of the ACM}.

\bibitem[{Munteanu and Sadownik(2019)}]{munteanu2019field}
Munteanu, C.; and Sadownik, S. 2019.
\newblock Field Studies of Interactive Technologies for Marginalized Users: A Canadian Ethics Policy Perspective.
\newblock \emph{Ageing and Digital Technology: Designing and Evaluating Emerging Technologies for Older Adults}.

\bibitem[{Müller, Sedley, and Ferrall-Nunge(2014)}]{muller2014survey}
Müller, H.; Sedley, A.; and Ferrall-Nunge, E. 2014.
\newblock \emph{Survey Research in HCI}.
\newblock Springer.
\newblock In J. Olson \& W. Kellogg (Eds.).

\bibitem[{Nazroo et~al.(2007)Nazroo, Jackson, Karlsen, and Torres}]{nazroo2007black}
Nazroo, J.; Jackson, J.; Karlsen, S.; and Torres, M. 2007.
\newblock The Black diaspora and health inequalities in the US and England: does where you go and how you get there make a difference?
\newblock \emph{Sociology of Health \& Illness}.

\bibitem[{Nicoletti and Bass(2023)}]{bloomberg2023humans}
Nicoletti, L.; and Bass, D. 2023.
\newblock Humans Are Biased. Generative AI Is Even Worse.

\bibitem[{Nierkens, de~Vries, and Stronks(2006)}]{nierkens2006smoking}
Nierkens, V.; de~Vries, H.; and Stronks, K. 2006.
\newblock Smoking in immigrants: do socioeconomic gradients follow the pattern expected from the tobacco epidemic?
\newblock \emph{Tobacco control}.

\bibitem[{{NIST}(2023)}]{standards2023ai}
{NIST}. 2023.
\newblock {AI Risk Management Framework}.

\bibitem[{Norton and Manson(1996)}]{norton1996research}
Norton, I.~M.; and Manson, S.~M. 1996.
\newblock Research in American Indian and Alaska Native communities: navigating the cultural universe of values and process.
\newblock \emph{Journal of consulting and clinical psychology}.

\bibitem[{Oldendick(2012)}]{oldendick2012survey}
Oldendick, R.~W. 2012.
\newblock Survey research ethics.
\newblock \emph{Handbook of survey methodology for the social sciences}.

\bibitem[{OpenAI(2023)}]{openai2023officialsurvey}
OpenAI. 2023.
\newblock Official ChatGPT survey - Shape the future of ChatGPT.

\bibitem[{Orimadegun(2020)}]{orimadegun2020protocol}
Orimadegun, A. 2020.
\newblock Protocol and Researcher’s Relationship with Institutional Review Board.
\newblock \emph{African Journal of Biomedical Research}.

\bibitem[{Ornstein(2013)}]{ornstein2013companion}
Ornstein, M. 2013.
\newblock \emph{A Companion to Survey Research}.
\newblock SAGE Publications.

\bibitem[{Othman(2023)}]{othman2023understanding}
Othman, K. 2023.
\newblock Understanding how moral decisions are affected by accidents of autonomous vehicles, prior knowledge, and perspective-taking: a continental analysis of a global survey.
\newblock \emph{AI and Ethics}.

\bibitem[{Palacios~Abad et~al.(2022)Palacios~Abad, Belding, Vigil-Hayes, and Zegura}]{palacios2022note}
Palacios~Abad, B.; Belding, E.; Vigil-Hayes, M.; and Zegura, E. 2022.
\newblock Note: Towards Community-Empowered Network Data Action.
\newblock In \emph{Proceedings of the 5th ACM SIGCAS/SIGCHI Conference on Computing and Sustainable Societies}.

\bibitem[{Patel et~al.(2013)Patel, Stevens, Puga et~al.}]{patel2013variations}
Patel, D.~I.; Stevens, K.~R.; Puga, F.; et~al. 2013.
\newblock Variations in institutional review board approval in the implementation of an improvement research study.
\newblock \emph{Nursing Research and Practice}.

\bibitem[{Paxton(2023)}]{paxton2023like}
Paxton, A. 2023.
\newblock What Is It Like to Sound Like a Bot?
\newblock \emph{Discourse and Writing/R{\'e}dactologie}.

\bibitem[{PE and MD(2019)}]{pe2019inconsistencies}
PE, E.; and MD, G. 2019.
\newblock Inconsistencies in institutional review board decisions: A proposal to regulate the decision-making process.
\newblock \emph{Bratislava Medical Journal/Bratislavske Lekarske Listy}.

\bibitem[{Peer et~al.(2022)Peer, Rothschild, Gordon, Evernden, and Damer}]{peer2022data}
Peer, E.; Rothschild, D.; Gordon, A.; Evernden, Z.; and Damer, E. 2022.
\newblock Data quality of platforms and panels for online behavioral research.
\newblock \emph{Behavior Research Methods}.

\bibitem[{Persson, Laaksoharju, and Koga(2021)}]{persson2021we}
Persson, A.; Laaksoharju, M.; and Koga, H. 2021.
\newblock We Mostly Think Alike: Individual Differences in Attitude Towards {AI} in Sweden and Japan.
\newblock \emph{The Review of Socionetwork Strategies}.

\bibitem[{Pfeffer et~al.(2022)Pfeffer, Mai, Weippl, Rader, and Krombholz}]{pfeffer2022replication}
Pfeffer, K.; Mai, A.; Weippl, E.; Rader, E.; and Krombholz, K. 2022.
\newblock Replication: Stories as Informal Lessons about Security.
\newblock In \emph{Eighteenth Symposium on Usable Privacy and Security (SOUPS 2022)}. USENIX Association.

\bibitem[{Phillips, Watkins, and Hammer(2018)}]{phillips2018beyond}
Phillips, A.~M.; Watkins, J.; and Hammer, D. 2018.
\newblock Beyond “asking questions”: Problematizing as a disciplinary activity.
\newblock \emph{Journal of Research in Science Teaching}.

\bibitem[{Posch et~al.(2018)Posch, Bleier, Fl{\"o}ck, Lechner, Kinder-Kurlanda, Helic, and Strohmaier}]{posch2018characterizing}
Posch, L.; Bleier, A.; Fl{\"o}ck, F.; Lechner, C.~M.; Kinder-Kurlanda, K.; Helic, D.; and Strohmaier, M. 2018.
\newblock Characterizing the global crowd workforce: A cross-country comparison of crowdworker demographics.
\newblock \emph{arXiv preprint arXiv:1812.05948}.

\bibitem[{Proctor and Schiebinger(2008)}]{proctor2008agnotology}
Proctor, R.~N.; and Schiebinger, L. 2008.
\newblock Agnotology: The making and unmaking of ignorance.

\bibitem[{Prolific(2023)}]{prolific2022}
Prolific. 2023.
\newblock Online participant recruitment for surveys and market research.

\bibitem[{Qualtrics(2023)}]{qualtrics2021}
Qualtrics. 2023.
\newblock Qualtrics XM - The Leading Experience Management Software.

\bibitem[{QueerInAI(2023)}]{queerinai2023queer}
QueerInAI, O.~O. 2023.
\newblock Queer In AI: A Case Study in Community-Led Participatory AI.
\newblock In \emph{Proceedings of the 2023 ACM Conference on Fairness, Accountability, and Transparency}. ACM.

\bibitem[{Rader, Wash, and Brooks(2012)}]{rader2012stories}
Rader, E.; Wash, R.; and Brooks, B. 2012.
\newblock Stories as Informal Lessons about Security.
\newblock In \emph{Proceedings of the Eighth Symposium on Usable Privacy and Security}. ACM.

\bibitem[{Rastogi et~al.(2023)Rastogi, Tulio~Ribeiro, King, Nori, and Amershi}]{rastogi2023supporting}
Rastogi, C.; Tulio~Ribeiro, M.; King, N.; Nori, H.; and Amershi, S. 2023.
\newblock Supporting Human-AI Collaboration in Auditing LLMs with LLMs.
\newblock In \emph{Proceedings of the 2023 AAAI/ACM Conference on AI, Ethics, and Society}. ACM.

\bibitem[{Rea and Parker(2014)}]{rea2014designing}
Rea, L.~M.; and Parker, R.~A. 2014.
\newblock \emph{Designing and conducting survey research: A comprehensive guide}.
\newblock John Wiley \& Sons.

\bibitem[{Reiser et~al.(2017)Reiser, Brody, Novak, Tipton, and Adams}]{reiser2017asking}
Reiser, B.~J.; Brody, L.; Novak, M.; Tipton, K.; and Adams, L. 2017.
\newblock Asking questions.
\newblock \emph{Helping students make sense of the world using next generation science and engineering practices}.

\bibitem[{Reyes(2019)}]{reyes2019eugenic}
Reyes, A. 2019.
\newblock Eugenic Visuality: Racist Epistemologies from Galton to ``The Bell Curve''.
\newblock \emph{Amerikastudien/American Studies}.

\bibitem[{Ribeiro et~al.(2019)Ribeiro, Saha, Babaei, Henrique, Messias, Benevenuto, Goga, Gummadi, and Redmiles}]{ribeiro2019microtargeting}
Ribeiro, F.~N.; Saha, K.; Babaei, M.; Henrique, L.; Messias, J.; Benevenuto, F.; Goga, O.; Gummadi, K.~P.; and Redmiles, E.~M. 2019.
\newblock On Microtargeting Socially Divisive Ads: A Case Study of Russia-Linked Ad Campaigns on Facebook.
\newblock In \emph{Proceedings of the {Conference} on {Fairness}, {Accountability}, and {Transparency}}. ACM.

\bibitem[{Rismani et~al.(2023)Rismani, Shelby, Smart, Jatho, Kroll, Moon, and Rostamzadeh}]{rismani2023plane}
Rismani, S.; Shelby, R.; Smart, A.; Jatho, E.; Kroll, J.; Moon, A.; and Rostamzadeh, N. 2023.
\newblock From Plane Crashes to Algorithmic Harm: Applicability of Safety Engineering Frameworks for Responsible ML.
\newblock In \emph{Proceedings of the 2023 {CHI} {Conference} on {Human} {Factors} in {Computing} {Systems}}. ACM.

\bibitem[{Roberts(2012)}]{roberts2012fatal}
Roberts, D. 2012.
\newblock \emph{Fatal Invention}.

\bibitem[{Rossi, Wright, and Anderson(2013)}]{rossi2013handbook}
Rossi, P.; Wright, J.; and Anderson, A. 2013.
\newblock \emph{Handbook of Survey Research}.
\newblock Elsevier Science.

\bibitem[{Rostamzadeh et~al.(2022)Rostamzadeh, Mincu, Roy, Smart, Wilcox, Pushkarna, Schrouff, Amironesei, Moorosi, and Heller}]{Healthsheet}
Rostamzadeh, N.; Mincu, D.; Roy, S.; Smart, A.; Wilcox, L.; Pushkarna, M.; Schrouff, J.; Amironesei, R.; Moorosi, N.; and Heller, K. 2022.
\newblock Healthsheet: Development of a Transparency Artifact for Health Datasets.
\newblock In \emph{Proceedings of the 2022 ACM Conference on Fairness, Accountability, and Transparency}. Association for Computing Machinery.

\bibitem[{Rust and Golombok(2014)}]{rust2014modern}
Rust, J.; and Golombok, S. 2014.
\newblock \emph{Modern psychometrics: The science of psychological assessment}.
\newblock Routledge.

\bibitem[{Said et~al.(2023)Said, Potinteu, Brich, Buder, Schumm, and Huff}]{said2023artificial}
Said, N.; Potinteu, A.~E.; Brich, I.; Buder, J.; Schumm, H.; and Huff, M. 2023.
\newblock An artificial intelligence perspective: How knowledge and confidence shape risk and benefit perception.
\newblock \emph{Computers in Human Behavior}.

\bibitem[{Sambasivan et~al.(2021)Sambasivan, Kapania, Highfill, Akrong, Paritosh, and Aroyo}]{sambasivan2021everyone}
Sambasivan, N.; Kapania, S.; Highfill, H.; Akrong, D.; Paritosh, P.; and Aroyo, L.~M. 2021.
\newblock ``Everyone Wants to Do the Model Work, Not the Data Work'': Data Cascades in High-Stakes AI.
\newblock In \emph{Proceedings of the 2021 {CHI} {Conference} on {Human} {Factors} in {Computing} {Systems}}. ACM.

\bibitem[{Schaeffer and Presser(2003)}]{schaeffer2003science}
Schaeffer, N.~C.; and Presser, S. 2003.
\newblock The science of asking questions.
\newblock \emph{Annual review of sociology}.

\bibitem[{Scharff et~al.(2010)Scharff, Mathews, Jackson, Hoffsuemmer, Martin, and Edwards}]{scharff2010more}
Scharff, D.~P.; Mathews, K.~J.; Jackson, P.; Hoffsuemmer, J.; Martin, E.; and Edwards, D. 2010.
\newblock More than Tuskegee: understanding mistrust about research participation.
\newblock \emph{Journal of health care for the poor and underserved}.

\bibitem[{Scharowski et~al.(2023)Scharowski, Benk, K\"{u}hne, Wettstein, and Br\"{u}hlmann}]{scharowski2023certification}
Scharowski, N.; Benk, M.; K\"{u}hne, S.~J.; Wettstein, L.; and Br\"{u}hlmann, F. 2023.
\newblock Certification Labels for Trustworthy AI: Insights From an Empirical Mixed-Method Study.
\newblock In \emph{Proceedings of the 2023 ACM Conference on Fairness, Accountability, and Transparency}. Association for Computing Machinery.

\bibitem[{Schrag(2010)}]{schrag2010ethical}
Schrag, Z.~M. 2010.
\newblock \emph{Ethical imperialism: Institutional review boards and the social sciences, 1965--2009}.
\newblock JHU Press.

\bibitem[{Schuler and Namioka(1993)}]{schuler1993participatory}
Schuler, D.; and Namioka, A. 1993.
\newblock \emph{Participatory design: Principles and practices}.
\newblock CRC Press.

\bibitem[{Schulz et~al.(2005)Schulz, Zenk, Kannan, Israel, Koch, and Stokes}]{schulz2005cbpr}
Schulz, A.~J.; Zenk, S.~N.; Kannan, S.; Israel, B.~A.; Koch, M.~A.; and Stokes, C.~A. 2005.
\newblock CBPR approaches to survey design and implementation.
\newblock \emph{Methods communitybased Particip Res Heal San Fr JosseyBass}.

\bibitem[{Science and Council(2023)}]{nataird2023strategic}
Science, N.; and Council, T. 2023.
\newblock National Artificial Intelligence Research and Development Strategic Plan, 2023 Update.

\bibitem[{Selwyn et~al.(2020)Selwyn, Cordoba, Andrejevic, and Campbell}]{selwyn2020ai}
Selwyn, N.; Cordoba, B.~G.; Andrejevic, M.; and Campbell, L. 2020.
\newblock AI for social good: Australian public attitudes toward AI and society.

\bibitem[{Septiandri et~al.(2023)Septiandri, Constantinides, Tahaei, and Quercia}]{septiandri2023weird}
Septiandri, A.~A.; Constantinides, M.; Tahaei, M.; and Quercia, D. 2023.
\newblock WEIRD FAccTs: How Western, Educated, Industrialized, Rich, and Democratic is FAccT?
\newblock In \emph{Proceedings of the 2023 ACM Conference on Fairness, Accountability, and Transparency}. ACM.

\bibitem[{Simonsen and Robertson(2012)}]{simonsen2012routledge}
Simonsen, J.; and Robertson, T. 2012.
\newblock \emph{Routledge international handbook of participatory design}.
\newblock Routledge.

\bibitem[{Sindermann et~al.(2021)Sindermann, Sha, Zhou, Wernicke, Schmitt, Li, Sariyska, Stavrou, Becker, and Montag}]{sindermann2021assessing}
Sindermann, C.; Sha, P.; Zhou, M.; Wernicke, J.; Schmitt, H.~S.; Li, M.; Sariyska, R.; Stavrou, M.; Becker, B.; and Montag, C. 2021.
\newblock Assessing the Attitude Towards Artificial Intelligence: Introduction of a Short Measure in German, Chinese, and English Language.
\newblock \emph{KI - K{\"u}nstliche Intelligenz}.

\bibitem[{Singer and Bishop(2021)}]{singer2021trust}
Singer, A.; and Bishop, M. 2021.
\newblock Trust-Based Security; Or, Trust Considered Harmful.
\newblock In \emph{Proceedings of the New Security Paradigms Workshop 2020}. ACM.

\bibitem[{Sinnott-Armstrong(2023)}]{consequentialism2023}
Sinnott-Armstrong, W. 2023.
\newblock Consequentialism.
\newblock In Zalta, E.~N.; and Nodelman, U., eds., \emph{The {Stanford} Encyclopedia of Philosophy}. Metaphysics Research Lab, Stanford University, {W}inter 2023 edition.

\bibitem[{Sloane et~al.(2022)Sloane, Moss, Awomolo, and Forlano}]{sloan2022participation}
Sloane, M.; Moss, E.; Awomolo, O.; and Forlano, L. 2022.
\newblock Participation Is Not a Design Fix for Machine Learning.
\newblock In \emph{Proceedings of the 2nd ACM Conference on Equity and Access in Algorithms, Mechanisms, and Optimization}. ACM.

\bibitem[{Smith, Christopher, and McCormick(2004)}]{smith2004development}
Smith, A.; Christopher, S.; and McCormick, A. K. H.~G. 2004.
\newblock Development and implementation of a culturally sensitive cervical health survey: a community-based participatory approach.
\newblock \emph{Women \& health}.

\bibitem[{Smith(2021)}]{smith2021decolonizing}
Smith, L.~T. 2021.
\newblock \emph{Decolonizing methodologies: Research and indigenous peoples}.
\newblock Bloomsbury Publishing.

\bibitem[{Soden, Toombs, and Thomas(2024)}]{soden2024evaluating}
Soden, R.; Toombs, A.; and Thomas, M. 2024.
\newblock Evaluating Interpretive Research in HCI.
\newblock \emph{Interactions}.

\bibitem[{Spector(2013)}]{spector2013survey}
Spector, P.~E. 2013.
\newblock Survey design and measure development.
\newblock \emph{The Oxford handbook of quantitative methods}.

\bibitem[{Srinivasan et~al.(2021)Srinivasan, Denton, Famularo, Rostamzadeh, Diaz, and Coleman}]{Artsheets}
Srinivasan, R.; Denton, E.; Famularo, J.; Rostamzadeh, N.; Diaz, F.; and Coleman, B. 2021.
\newblock Artsheets for art datasets.
\newblock In \emph{Thirty-fifth conference on neural information processing systems datasets and benchmarks track (round 2)}.

\bibitem[{Sunarti et~al.(2021)Sunarti, Rahman, Naufal, Risky, Febriyanto, and Masnina}]{sunarti2021artificial}
Sunarti, S.; Rahman, F.~F.; Naufal, M.; Risky, M.; Febriyanto, K.; and Masnina, R. 2021.
\newblock Artificial intelligence in healthcare: opportunities and risk for future.
\newblock \emph{Gaceta Sanitaria}.

\bibitem[{SurveyMonkey(2024)}]{surveymonkey2024calculate}
SurveyMonkey. 2024.
\newblock Calculate your sample size.

\bibitem[{Tahaei et~al.(2023)Tahaei, Constantinides, Quercia, and Muller}]{tahaei2023systematic}
Tahaei, M.; Constantinides, M.; Quercia, D.; and Muller, M. 2023.
\newblock A Systematic Literature Review of Human-Centered, Ethical, and Responsible AI.

\bibitem[{Tan and Cabato(2023)}]{tan2023behind}
Tan, R.; and Cabato, R. 2023.
\newblock Behind the AI boom, an army of overseas workers in ‘digital sweatshops’.

\bibitem[{Tang, Birrell, and Lerner(2022)}]{tang2022replication}
Tang, J.; Birrell, E.; and Lerner, A. 2022.
\newblock Replication: How Well Do My Results Generalize Now? The External Validity of Online Privacy and Security Surveys.
\newblock In \emph{Eighteenth Symposium on Usable Privacy and Security (SOUPS 2022)}. USENIX Association.

\bibitem[{Tapaha(2017)}]{tapaha2017we}
Tapaha, O.~G. 2017.
\newblock " We Lived It": Stories of Cultural Resilience, Din{\'e}k’ehgo Nanitiin (Din{\'e}-Based Instruction), and Navigating between University and Tribal Institutional Review Boards.

\bibitem[{Tillyard and DeGennaro~Jr(2019)}]{tillyard2019new}
Tillyard, G.; and DeGennaro~Jr, V. 2019.
\newblock New methodologies for global health research: Improving the knowledge, attitude, and practice survey model through participatory research in Haiti.
\newblock \emph{Qualitative Health Research}.

\bibitem[{Torkamaan et~al.(2024)Torkamaan, Tahaei, Buijsman, Xiao, Wilkinson, and Knijnenburg}]{Torkamaan2024}
Torkamaan, H.; Tahaei, M.; Buijsman, S.; Xiao, Z.; Wilkinson, D.; and Knijnenburg, B.~P. 2024.
\newblock \emph{The Role of Human-Centered AI in User Modeling, Adaptation, and Personalization---Models, Frameworks, and Paradigms}.
\newblock Springer Nature Switzerland.

\bibitem[{Tourangeau, Rips, and Rasinski(2000)}]{tourangeau2000psychology}
Tourangeau, R.; Rips, L.~J.; and Rasinski, K. 2000.
\newblock The psychology of survey response.

\bibitem[{Tourangeau and Smith(1996)}]{tourangeau1996asking}
Tourangeau, R.; and Smith, T.~W. 1996.
\newblock Asking sensitive questions: The impact of data collection mode, question format, and question context.
\newblock \emph{Public opinion quarterly}.

\bibitem[{Trefzer et~al.(2014)Trefzer, Jackson, McKee, and Dellinger}]{trefzer2014introduction}
Trefzer, A.; Jackson, J.~T.; McKee, K.; and Dellinger, K. 2014.
\newblock Introduction: The global south and/in the global north: Interdisciplinary investigations.
\newblock \emph{The Global South}.

\bibitem[{{United Nations}(2022)}]{un2024AI}
{United Nations}. 2022.
\newblock A Future with AI: Voices of Global Youth Report Launched.

\bibitem[{van Berkel, Sarsenbayeva, and Goncalves(2023)}]{van_berkel2023methodology}
van Berkel, N.; Sarsenbayeva, Z.; and Goncalves, J. 2023.
\newblock The methodology of studying fairness perceptions in Artificial Intelligence: Contrasting CHI and FAccT.
\newblock \emph{International Journal of Human-Computer Studies}.

\bibitem[{Veselovsky, Ribeiro, and West(2023)}]{veselovsky2023artificial}
Veselovsky, V.; Ribeiro, M.~H.; and West, R. 2023.
\newblock Artificial Artificial Artificial Intelligence: Crowd Workers Widely Use Large Language Models for Text Production Tasks.

\bibitem[{Wacharamanotham et~al.(2020)Wacharamanotham, Eisenring, Haroz, and Echtler}]{wacharamanotham2020transparency}
Wacharamanotham, C.; Eisenring, L.; Haroz, S.; and Echtler, F. 2020.
\newblock Transparency of CHI Research Artifacts: Results of a Self-Reported Survey.
\newblock In \emph{Proceedings of the 2020 CHI Conference on Human Factors in Computing Systems}. ACM.

\bibitem[{Wang, Prabhat, and Sambasivan(2022)}]{wang2022whose}
Wang, D.; Prabhat, S.; and Sambasivan, N. 2022.
\newblock Whose AI Dream? In Search of the Aspiration in Data Annotation.
\newblock In \emph{Proceedings of the 2022 {CHI} {Conference} on {Human} {Factors} in {Computing} {Systems}}. ACM.

\bibitem[{Wilcox, Brewer, and Diaz(2023)}]{wilcox2023aiconsent}
Wilcox, L.; Brewer, R.; and Diaz, F. 2023.
\newblock AI Consent Futures: A Case Study on Voice Data Collection with Clinicians.
\newblock \emph{Proc. ACM Hum.-Comput. Interact.}

\bibitem[{Wilcox et~al.(2023)Wilcox, Shelby, Veeraraghavan, Haimson, Erickson, Turken, and Gulotta}]{wilcox2023infra}
Wilcox, L.; Shelby, R.; Veeraraghavan, R.; Haimson, O.~L.; Erickson, G.~C.; Turken, M.; and Gulotta, R. 2023.
\newblock Infrastructuring Care: How Trans and Non-Binary People Meet Health and Well-Being Needs through Technology.
\newblock In \emph{Proceedings of the 2023 CHI Conference on Human Factors in Computing Systems}. Association for Computing Machinery.

\bibitem[{Winston(2020)}]{winston2020scientific}
Winston, A.~S. 2020.
\newblock Scientific racism and North American psychology.
\newblock In \emph{Oxford research encyclopedia of psychology}.

\bibitem[{Wong and Khovanskaya(2018)}]{wong2018speculative}
Wong, R.~Y.; and Khovanskaya, V. 2018.
\newblock \emph{Speculative design in HCI: from corporate imaginations to critical orientations}.
\newblock Springer.

\bibitem[{Yigitcanlar et~al.(2020)Yigitcanlar, Desouza, Butler, and Roozkhosh}]{yigitcanlar2020contributions}
Yigitcanlar, T.; Desouza, K.~C.; Butler, L.; and Roozkhosh, F. 2020.
\newblock Contributions and Risks of Artificial Intelligence (AI) in Building Smarter Cities: Insights from a Systematic Review of the Literature.
\newblock \emph{Energies}.

\bibitem[{Young, Katell, and Krafft(2022)}]{young2022confronting}
Young, M.; Katell, M.; and Krafft, P. 2022.
\newblock Confronting Power and Corporate Capture at the FAccT Conference.
\newblock In \emph{2022 {ACM} {Conference} on {Fairness}, {Accountability}, and {Transparency}}. ACM.

\bibitem[{Zanetti, Iseppi, and Cassese(2019)}]{zanetti2019psychopathic}
Zanetti, M.; Iseppi, G.; and Cassese, F.~P. 2019.
\newblock A ``psychopathic'' Artificial Intelligence: the possible risks of a deviating AI in Education.
\newblock \emph{Research on Education and Media}.

\bibitem[{Zhang and Dafoe(2020)}]{zhang2020us}
Zhang, B.; and Dafoe, A. 2020.
\newblock U.S. Public Opinion on the Governance of Artificial Intelligence.
\newblock In \emph{Proceedings of the AAAI/ACM Conference on AI, Ethics, and Society}. ACM.

\end{thebibliography}

\appendix


\section{Known Limitations of Surveys}
\label{app:known-issues}
There are several well-known issues with surveys that are often acknowledged as limitations in studies. Below, we list several common biases; however, this is not an exhaustive list. For a more comprehensive discussion, we refer to survey design papers and textbooks such as \cite{muller2014survey, krosnick1999survey, choi2005catalog}.

\begin{itemize}
    \item \textbf{Acquiescence bias and experimenter effect}: Survey respondents may agree with a question or statement regardless of their actual feelings or attitudes. This can stem from a desire to be agreeable, a lack of motivation in answering questions, due to the influence of the researcher or the institution conducting the survey, or due to desire to satisfy what respondents think the researchers' expectations from the study are.
    \item \textbf{Satisficing:} Respondents may opt for answers that merely satisfy the survey's requirements, rather than seeking the most accurate or optimal response. This behavior could be due to the effort involved, distractions, or a lack of interest in the survey's outcomes.
    \item \textbf{Social desirability:} In response to sensitive questions, survey participants may answer in a manner they believe will be viewed favorably by others. For instance, questions about sexual orientation or taboo subjects might elicit responses that do not accurately reflect the respondent's true stance.
    \item \textbf{Question and response order bias:} The sequencing of questions or answer options can influence survey results. The order in which questions are presented can prime respondents' views as they progress through the survey. Similarly, non-randomized answer choices may not have an equal chance of being selected.
    \item \textbf{Framing effects:} The choice of wording in questions may affect survey responses. For example, users' agreement or disagreement with specific statements may depend on whether those statements are positively or negatively framed. The choice of specific words (even among synonyms, such as concern vs. worry vs. fear vs. discomfort) may also lead to different outcomes. Nuances in translation may further complicate the interpretation of questions and answers in multi-lingual studies. Other wording choices and associated mistakes (e.g., double negation, double-barreled questions, leading questions, etc.) can decrease respondents' comprehension of the questions and introduce biases in the analysis, compromising the objectivity and accuracy of the survey results. More broadly, framing effects may also emerge from the overall narrative of the survey, for example, suggesting a dichotomous trade-off between benefits and risks, without considering other nuances of the discourse or other potential factors affecting respondents' perspectives. 
    \item \textbf{Sampling bias:}
    Collecting survey responses from a non-representative sample reduces the generalizability of the results to the broader target population, which may differ in socio-demographics and experiences. Similarly, by focusing on AI users, researchers exclude the perspectives of those who do not use AI or lack the expertise, experience, motivation, or resources to use it. Conducting research in a single country or language further limits the ability to capture cross-cultural differences. Even the data collection format (e.g., online vs. pen-and-paper surveys, accessibility features) can impact results by excluding individuals who lack the necessary knowledge, skills, access, or physical or cognitive abilities to complete the survey\end{itemize}

Despite the acknowledged limitations and known issues, surveys remain a frequently employed method for gathering data on opinions, behaviors, attitudes, knowledge, personal characteristics, and motivations. This trend extends to AI research as well, with surveys playing a significant role in this field (for examples, refer to Section~\ref{sec:rw-survey-ai-research}).

\section{Additional Materials for Pilot Survey} 
\label{app:survey-additional}

\subsection{Survey Instrument}
\label{app:survey}
[Answer options to close-ended questions were randomized where appropriate.]
\begin{itemize}
    \item Have you heard of the term ``Artificial Intelligence'' (or ``AI'')?
    
    $-$ Yes, I have heard of the term ``Artificial Intelligence'' (or ``AI''), and I feel confident explaining what it means to an expert.
    
    $-$ Yes, I have heard of the term ``Artificial Intelligence'' (or ``AI''). However, I do not feel confident explaining what it means to an expert.
    
    $-$ No, I have not heard of the term ``Artificial Intelligence'' (or ``AI'').\\
    
    The following questions are based on your understanding of \textit{existing} AI systems.
    
    \item How do you think existing AI systems could \textit{benefit} you? Please give details (at least 100 characters).

    \item We want to know what you have learned from others about the \textit{benefits} of \textit{existing} AI systems. Specifically, we are interested in stories you have heard about the benefits of existing AI systems from \textit{OTHER PEOPLE}, such as friends, coworkers, social media sites, TV shows, news websites, blogs, or any other sources---NOT experiences that happened to you personally. Describe in detail the most memorable story (at least 100 characters).

    \item How did you hear about that story?\\

    \item 
    How do you think existing AI systems could put you \textit{at risk}? Please give details (at least 100 characters).

    \item We want to know what you have learned from others about the \textit{risks} of \textit{existing} AI systems. Specifically, we are interested in stories you have heard about the risks of existing AI systems from \textit{OTHER PEOPLE}, such as friends, coworkers, social media sites, TV shows, news websites, blogs, or any other sources---NOT experiences that happened to you personally. Describe in detail the most memorable story (at least 100 characters).

    \item How did you hear about that story?\\
    
    \item Please select Excellent to show you are paying attention to this question [Very Poor, Poor, Fair, Good, Excellent].\\
    
    In this section, we want to learn about the different ways you \textit{envision} AI systems working.
    \item If you had a magic wand that could create an AI system, what would you want that AI system to do for you? Please give details (at least 100 characters).

    \item How could the AI system that you just described put you (or someone else) \textit{at risk}? Please give details (at least 100 characters).
    
    Please answer the question below given the following definition of an AI system:\\ \textit{``An AI system is a technology that can generate outputs such as predictions, recommendations, or decisions influencing real or online spaces. AI systems are designed to work with different levels of independence, meaning some might need more human guidance, while others can handle tasks on their own.''}
    
    \item Based on this definition, in your opinion, what characteristics should an AI system have to be \textit{trustworthy}? Please describe ``your'' understanding using your own words. Please give details (at least 100 characters).\\
    \item Please select Rarely to show you are paying attention to this question [Always, Never, Rarely].\\
    
    [Demographics]
    \item What best describes your employment status over the last three months?
    
    $-$ Working full-time
    
    $-$ Working part-time
    
    $-$ Unemployed and looking for work
    
    $-$ A homemaker or stay-at-home parent 
    
    $-$ Student
    
    $-$ Retired
    
    $-$ Other
    
    \item What ethnic group describes you the best? [Open-ended]
    
    \item What is the highest level of education you have completed?
    
    $-$ Some high school or less
    
    $-$ High school diploma or GED
    
    $-$ Some college, but no degree
    
    $-$ Associates or technical degree
    
    $-$ Bachelor’s degree
    
    $-$ Graduate or professional degree (MA, MS, MBA, PhD, JD, MD, DDS, etc.) 
    
    $-$ Prefer not to say

    \item Please select the option that best describes your personal income relative to others in your age group and location.
    
    $-$ Below average 
    
    $-$ Average 
    
    $-$ Above average 
    
    $-$ Unsure
    
    $-$ Prefer not to say
    
    \item Please answer the following questions [Yes, Sort of, No].
    
    $-$ I know how to program in at least one programming language.
    
    $-$ My family members or friends often ask me for computing-related advice.
    
    $-$ I study or work in IT or a computing-related field.
    
\end{itemize}


\subsection{Participant Demographics}
\label{app:demographics}

\begin{table}[ht]
\caption{Participant demographics (N=282).}
\label{tab:demographics}
\begin{tabular}{@{}lr@{}}
\toprule
\textit{Country of residence} & \\
\quad Australia & 50 (18\%) \\
\quad Israel & 48 (17\%) \\
\quad Chile & 47 (17\%) \\
\quad United Kingdom & 47 (17\%) \\
\quad United States & 46 (16\%) \\
\quad South Africa & 44 (16\%) \\
\textit{Ethnicity} &  \\
\quad White & 161 (57\%) \\
\quad Black & 48 (17\%) \\
\quad Mixed & 34 (12\%) \\
\quad Other & 19 (7\%) \\
\quad Asian & 18 (6\%) \\
\quad Not available & 2 (1\%) \\
\textit{Employment status} &  \\
\quad Working full-time & 126 (45\%) \\
\quad Working part-time & 54 (19\%) \\
\quad Student & 54 (19\%) \\
\quad Unemployed and looking for work & 34 (12\%) \\
\quad Other & 5 (2\%) \\
\quad Retired & 5 (2\%) \\
\quad A homemaker or stay-at-home parent & 4 (1\%) \\
\textit{Gender} & \\
\quad Female & 143 (51\%) \\
\quad Male & 139 (49\%) \\
\textit{Income relative to age group and location} &  \\
\quad Average & 116 (41\%) \\
\quad Below average & 79 (28\%) \\
\quad Above average & 66 (23\%) \\
\quad Prefer not to say & 11 (4\%) \\
\quad Unsure & 10 (4\%) \\
\textit{Familiarity with AI} &  \\
\quad Heard but can't explain to an expert & 169 (60\%) \\
\quad Heard and can explain to an expert & 113 (40\%) \\
Technical background & 62 (22\%) \\ \bottomrule
\end{tabular}%
\end{table}

The final dataset consisted of participants from six countries.  Table~\ref{tab:demographics} summarizes the demographics of our participants. We achieved a balanced sample in terms of gender. 
Racially and ethnically, many participants 
(57\%) identified as White. 
In terms of employment status, several participants 
were employed full-time (45\%), and 
described their income as average (41\%). 

Concerning familiarity with AI,~169 participants (60\%) expressed that they had heard of the term but did not feel confident explaining its meaning to an expert, and~113 participants (40\%) were familiar with AI and felt confident explaining its meaning to an expert. No participant claimed to be unfamiliar with the term. Based on our criteria for assessing the technical background of participants, ~62 participants (22\%) had some technical background.

\subsection{Reflections: Impact of Researcher Tools, Practices, and Choices}
\label{app:reflections}
In this section, we revisit the decisions made during the design, deployment, and analysis of our pilot survey that are not necessarily covered in the literature review but were crucial considerations. Each subsection addresses a specific question we encountered and had to deliberate upon. A more concise set of heuristic questions can be found in Section~\ref{sec:value-tensions}.

\subsubsection{How to Frame Questions?}
Our pilot survey was rooted in prior research with public perception of AI and echoed their consequentialist ethics framing---``\textit{the view that normative properties depend only on consequences}''~\cite{consequentialism2023}. In our case, the framing of the questions with ``How do you think existing AI systems could \textit{benefit} you?'' explicitly emphasizes the consequences instead of anything else. From an alternative standpoint such as deontologist---``\textit{normative theories regarding which choices are morally required, forbidden, or permitted . . . In contrast to consequentialist theories, deontological theories judge the morality of choices by criteria different from the states of affairs those choices bring about}''~\cite{sep-ethics-deontological}---, we could have instead asked about what rules should AI models or systems follow or not follow, instead of asking about specific desired and risky consequences: ``What rules should a beneficial AI technology follow?'' Alternatively, we could have focused on virtue ethics---``\textit{. . . the one that emphasizes the virtues, or moral character, in contrast to the approach that emphasizes duties or rules (deontology) or that emphasizes the consequences of actions (consequentialism)}''~\cite{sep-ethics-virtue}---and asked more about the traits or virtues that AI should have: ``How would you describe a generous AI?'' 

As for deployment of terms such as ``risk'' and ``benefit'' in the survey, it is in fact imperative to understand, in a much more localized way, the unique cultural and philosophical underpinnings of concepts like ``risk'' and ``benefit'' in order to write reliable survey questions on these concepts, given their cultural meaning and situation, and the high degree of influence that their surrounding social values, norms, and beliefs play in people's actual risk assessments~\cite{douglas1983risk,boholm1996risk, beck1992risk, bernstein1996against,beck1996world}.

All these viewpoints are contingent on definitions that vary cross-culturally. Without transparent reporting and the availability of artifacts, readers are unable to determine precisely what questions were asked, making it impossible to know whether there was a downplaying or overlooking of potential concerns raised by participants. Participants may express concerns or exhibit excitement, but this does not necessarily mean that their concerns are outweighed by the perceived benefits, especially when considering the actual net benefit or harm to others. However, results are sometimes simplistically reported as statements like ``Global Study Shows Optimism About AI's Potential''~\cite{google2024global} or ``AI is making the world more nervous,''~\cite{boyon2023nervous}, failing to capture these nuances and trade-offs.

\subsubsection{Should We Translate the Survey?}
The practice of translating questions from English into other languages for comparative analysis warrants attention. While such translations, which may be conducted meticulously and carefully~\cite[e.g., ][]{kelley2021exciting}, aim to include more countries or cultures, they may not accurately reflect other cultures' perspectives. Concluding that direct experiences with AI alone drive excitement and alleviate concerns is an oversimplification. Attitudes toward AI are shaped by various factors, including media representation, personal experiences and beliefs, and societal narratives, and not solely by direct interactions with AI. Individuals may lack meta-awareness of these other factors influencing their perceptions. Direct interactions with AI could be detrimental to some individuals, and those most vulnerable to being marginalized by AI may be individuals who have not experienced it firsthand. Furthermore, survey results indicating a pronounced positive orientation in certain regions and emerging markets~\cite[e.g., ][]{google2024global} could result from different stages of AI adoption, economic factors, or cultural attitudes toward technology, which may not be adequately captured in a survey.

We considered using AI for (1) generating translations of questions, (2) translating participant responses, and (3) analyzing responses. However, given the ongoing debates about employing Large Language Models (LLMs) in research~\cite[e.g., ][]{paxton2023like, rastogi2023supporting, hosseini2023ethics, byun2023dispensing}, we realized that using AI without informing participants would be inappropriate. We took into account the following: (1) the risk that participants' data might be absorbed into corporate-owned LLMs, creating a permanent trace of their participation and ideas, (2) the accuracy of AI in conveying participants' true intentions and responses, and (3) the potential use of these responses for future model training, and the impact of over-surveyed populations on these models.

In any of these use cases, we were uncertain about how to seek consent from our participants for using AI to analyze their data. Typically, consent for participation in research covers privacy, data security, and its use by the research team. Moving forward, however, researchers, including those in UX, may need to offer participants the option to have their data analyzed by AI, with full disclosure of what this entails. Under such an approach, would researchers need to provide participants with the analysis summary for their final approval? Should participants be given the chance to review the summary if AI is used, to ensure it aligns with their perspective, considering that the analysis is no longer conducted solely by individuals trained in positionality and research ethics?

\subsubsection{Where to Find Participants?}
Our team's diversity, spanning various affiliations, faced limitations in participant recruitment due to being constrained by choosing appropriate platforms. Differing experiences with platforms such as MTurk and Prolific among team members added to these challenges. For instance, institutional policies prevented one member from using Prolific, despite their willingness to fund the project. Consequently, another team member, with more flexibility in platform choice and budget, took responsibility for project funding. The literature review also indicates a wide range of recruitment methods, from crowdsourcing platforms to emails and professional research companies.

Prolific's reach was limited in terms of global access. We attempted to recruit an equal number of male and female participants from over 30 countries, aiming for 50 participants from each country, but were unsuccessful in many countries. Consequently, our final participant pool comprised individuals from Australia, Chile, Israel, the United Kingdom, the United States, and South Africa. Notably, there was an unexpected lack of participation from regions such as China, India, South America, and Africa, some of the most populous areas globally. Additionally, Prolific's enforced binary gender option further limited our reach to populations that do not fall into the binary classification of gender.

Additionally, the use of online platforms excludes certain groups, such as those without Internet access, children, individuals in countries with different payment methods, and people not connected to platform users, resulting in a sampling bias. Prolific primarily recruits participants through word of mouth and social media.\footnote{``Participants are primarily recruited to Prolific via word of mouth, including word of mouth via social media. When Prolific was founded in 2014, our participants were recruited via three channels: 1) Social media (e.g., Facebook, Twitter, Reddit, and various other online forums). 2) Flyer distribution on university campuses. 3) The Prolific referral scheme (ceased March 2019). This allowed participants to invite their social network to join Prolific, in return for small cash incentives for the referrer''~\cite{prolific2022}.} While convenient, results may be skewed toward those with stronger opinions or experiences with AI, as people more interested in or affected by AI are more likely to participate, creating a self-selection bias. This bias is particularly evident in crowdsourcing platforms like Prolific or MTurk, where individuals can choose which tasks to take, with interest in the survey topic being a significant motivator for participation~\cite{kaufmann2011more}. However, prior work indicates that this is less of a prominent reason for \textit{professional} MTurk users, who primarily select surveys based on compensation rather than interest in the topic~\cite{lovett2018data}.

\subsubsection{How Much Should We Pay Participants?} In addition to access considerations, we decided to compensate all participants based on the payment rates of our home institution. However, this decision prompts a critical question: Should compensation be adjusted based on the participant's country or kept consistent across all countries? This dilemma underscores the challenges of fair participant treatment across diverse regions and economies, emphasizing the need for a nuanced compensation approach that considers both institutional standards and the participants' economic situations.

Examining the data from our literature review, it is evident that researchers employ diverse methods for compensating survey participants, ranging from unspecified (not included in the paper) to free or paid, with varying amounts reported (refer to Section~\ref{sec:lit-review-findings} for detailed findings). Given the well-documented ethical concerns surrounding data annotation practices in AI research, particularly in terms of labor and payment~\cite{wang2022whose, tahaei2023systematic, tan2023behind}, the question of what constitutes an ethical way of compensating survey participants remains open for discussion.

\subsubsection{How to Analyze Data?}
Finding a suitable framework for a top-down analysis of data, especially regarding the benefits and risks of AI, was difficult. This challenge arises mainly because most existing frameworks come from Western countries, and there is a lack of comprehensive frameworks or taxonomies for a complex and rapidly changing field like AI. For example, the difficulties in applying AI safety taxonomies are well-recognized due to the ever-evolving nature of AI~\cite{rismani2023plane}.

Given these constraints, our approach predominantly involved bottom-up analysis for most of the questions. An exception was made for the ``trustworthiness'' question, in which we applied a combination of top-down and bottom-up methods. For the top-down component, we used the NIST AI Risk Management Framework~\cite{standards2023ai} as a starting point. However, this choice introduced a US-centric bias into our analysis, potentially excluding or marginalizing other cultural views. The use of a framework developed within a specific cultural and institutional context raises important questions about the universality and applicability of our analysis across different contexts. Our positionality statement provides an initial reflection on our analytical perspective. Yet, a more in-depth consideration of the frameworks we use and their impact on our results highlights the broader issue of institutional and methodological positionalities. This presents a question for our research community: How can we effectively address these positionalities?

\section{Additional Materials for the Systematic Literature Review} 
\label{app:details-systematic}

\subsection{Query}
\label{app:query-review}

\begin{itemize}

\item \textit{Who? Public.} ``public'' OR ``representative'' OR ``population'' OR ``citizen'' OR ``citizens'' OR ``civic'' OR ``community'' OR ``non-expert'' OR ``non-experts''.
\item \textit{What? AI.} ``artificial intelligence'' OR ``machine learning'' OR ``deep learning'' OR ``AI''.
\item \textit{What? Perceptions.} ``thoughts'' OR ``feel'' OR ``feels'' OR ``feeling'' OR ``experience'' OR ``experiences'' OR ``feelings'' OR ``perception'' OR ``perceptions'' OR ``perceive'' OR ``attitude'' OR ``attitudes'' OR ``opinion'' OR ``opinions'' OR ``view'' OR ``views''.
\item \textit{How? Surveys.} ``survey'' OR ``surveys'' OR ``poll'' OR ``polls'' OR ``questionnaire'' OR ``questionnaires''.
\end{itemize}

\subsection{Prisma Diagram}
\label{app:prisma}

\begin{itemize}
    \item ACM DL ($n=45$): We searched the entire ACM database, applying the aforementioned query to titles or abstracts from the most recent two years. This yielded nine records, with two overlapping with FAccT and AIES searches, and one exclusion as it was not a survey of people. For FAccT ($n=18$) and AIES ($n=17$), we executed the query on FAccT and AIES papers searching by title or abstract without a date restriction to broaden our reach. We also excluded the AI clause, as these conferences are inherently centered on AI and ethics. After manually screening the results, we included five papers from FAccT and four from AIES in our final collection.
    \item AI \& Ethics ($n=92$) and AI \& Society ($n=126$): We carried out a full-text search in Springer’s database for the past two years using our query, noting that the database does not offer options for searching by title or abstract. All titles and abstracts were manually reviewed, applying our exclusion criteria, resulting in 26 papers from AI \& Society and six from AI \& Ethics being selected. Following a secondary review, three papers from AI \& Society were excluded, while all papers from AI \& Ethics were retained.
\end{itemize}

\begin{figure}
    \centering
    \includegraphics[width=.95\linewidth]{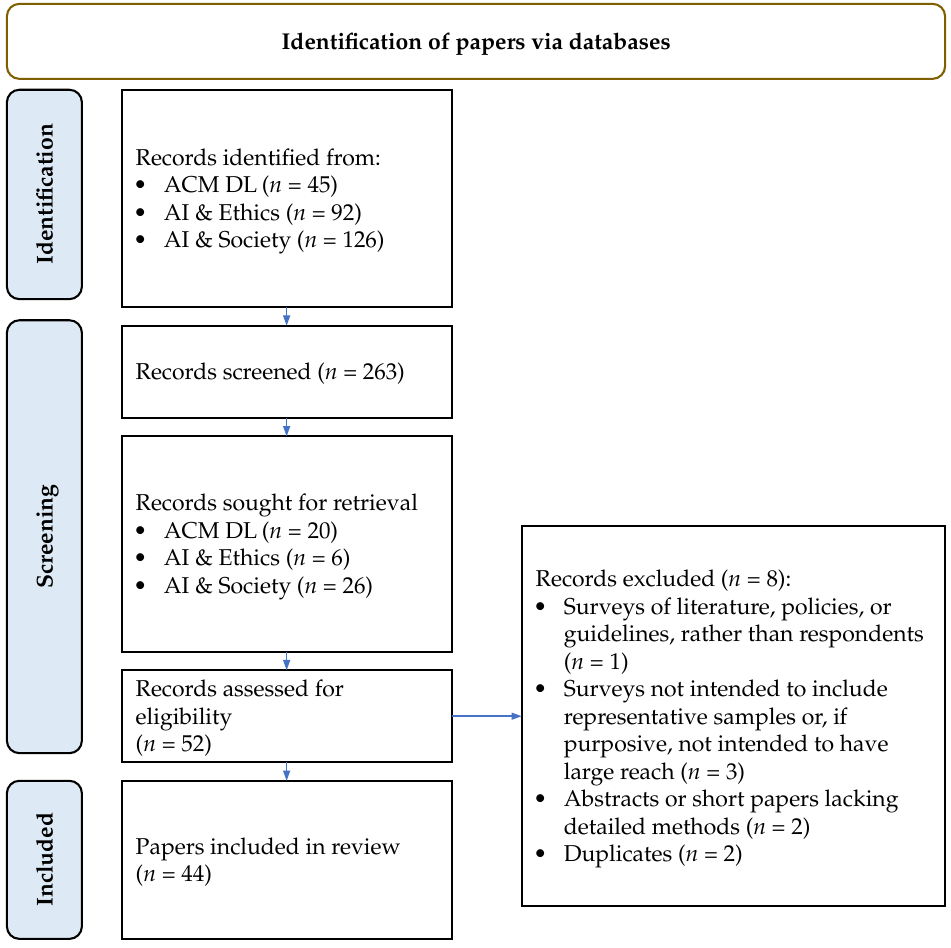}
    \caption{Prisma diagram for the systematic literature review. See Sections~\ref{sec:lit-review} and \ref{app:prisma} for details.}
    \label{fig:prisma}
\end{figure}

\end{document}